\documentclass[prd,aps,twocolumn,a4paper,showkeys,nofootinbib]{revtex4-1}

\usepackage{graphicx,psfrag}
\usepackage{mathrsfs}
\usepackage{amsmath,amsfonts,amssymb}
\usepackage{multirow}
\usepackage{comment}
\usepackage{ulem}
\usepackage{hyperref}
\usepackage{enumitem}
\usepackage{natbib}

\newcommand{\be}{\begin{equation}}
\newcommand{\ee}{\end{equation}}
\newcommand{\bea}{\begin{eqnarray}}
\newcommand{\eea}{\end{eqnarray}}

\def\lm{\ell m}

\def\Msun{{\rm M_{\odot}}}
\def\GPc3yr{{\rm Gpc}^{-3} {\rm yr^{-1}}}
\def\GMc2{{\rm G M_{\odot} c^{-2}}}

\def\B{\mathcal{B}}

\def\kt2{\kappa^\text{T}_2}

\def\Mbh{M_\text{BH}}
\def\abh{a_\text{BH}}
\def\Mns{M_\text{NS}}

\def\Mfbh{M_\bullet}
\def\afbh{a_\bullet}
\def\etal{{\it et al.}}

\usepackage{pifont} 

\newcommand{\teob}{\texttt{TEOBResumS}}
\newcommand{\TEOB}{\teob}
\newcommand{\bajes}{\texttt{bajes}}
\newcommand{\SEOB}[1]{\texttt{SEOBNR{#1}}}
\newcommand{\Phenom}[1]{\texttt{Phenom{#1}}}

\begin{document}

\title{Numerical-Relativity-Informed Effective-One-Body model for Black-Hole-Neutron-Star Mergers with Higher Modes and Spin Precession}

\author{Alejandra \surname{Gonzalez}$^{1}$, 
Rossella \surname{Gamba}$^{1}$,
Matteo \surname{Breschi}$^{1}$,\\
Francesco \surname{Zappa}$^{1}$,
Gregorio \surname{Carullo}$^{1}$,
Sebastiano \surname{Bernuzzi}$^{1}$, 
Alessandro \surname{Nagar}$^{2,3}$}

\affiliation{${}^1$Theoretisch-Physikalisches Institut, Friedrich-Schiller-Universit{\"a}t Jena, Fr{\"o}belstieg 1, 07743 Jena, Germany}
\affiliation{${}^2$INFN Sezione di Torino, Via P. Giuria 1, 10125 Torino, Italy}
\affiliation{${}^3$Institut des Hautes Etudes Scientifiques, 91440 Bures-sur-Yvette, France}

\date{\today}
\begin{abstract}
  We present the first effective-one-body (EOB) model for
  generic-spins quasi-circular black-hole--neutron-star (BHNS)
  inspiral-merger-ringdown gravitational waveforms (GWs). Our model is based on a new numerical-relativity
  (NR) informed expression of the BH remnant and its ringdown. It
  reproduces the NR $(\ell,m)=(2,2)$ waveform with
  typical phase agreement of ${\lesssim0.5}\,$rad (${\lesssim1}\,$rad) to merger (ringdown).
  The maximum (minimum) mismatch between the $(2,2)$ and the NR data
  is 4\% (0.6\%).
  Higher modes (HMs) $(2,1)$, $(3,2)$, $(3,3)$, $(4,4)$ and $(5,5)$ are included and their
  mismatch with the available NR waveforms are up to (down to) a 60\% (1\%) depending on the inclination.
  Phase comparison with a 16 orbit precessing simulation shows differences within the NR uncertainties.  
  We demonstrate the applicability of the model in GW parameter
  estimation by perfoming the first BHNS Bayesian analysis with HMs
  (and non-precessing spins) of the event GW190814, together with
   new $(2,2)$-mode analysis of GW200105 and GW200115. 
  For the GW190814 study, the inclusion of HMs gives tighter parameter posteriors.
  The Bayes factors of our analyses on this event show decisive evidence
  for the presence of HMs, but no clear preference for a BHNS or a
  binary black hole (BBH) source. 
  Similarly, we confirm GW200105 and GW200115 show no evidence for tidal effects.
\end{abstract}

\pacs{
  04.25.D-,     
  04.30.Db,   
  95.30.Sf,     
  95.30.Lz,   
  97.60.Jd      
}

\maketitle

\section{Introduction} 
The recent gravitational wave (GW) observations GW200105 and GW200115 can 
be interpreted as the first detections of black-hole--neutron-star (BHNS) mergers~\cite{LIGOScientific:2021qlt}. These observations allow to estimate a merger rate of $45\,\GPc3yr$, 
for masses consistent with the ones of these observed signals, and $130\,\GPc3yr$ for a broader BHNS mass distribution~\cite{LIGOScientific:2021qlt}. 
For typical neutron star masses $1.4\,\Msun\lesssim \Mns \lesssim 3\,\Msun$~\cite{PhysRevLett.32.324,Godzieba:2020tjn} and masses from currently observed populations of binary black holes 
(BBH), BHNS are expected to have mass ratios, $q=m_1/m_2$ ($m_1 \geq m_2$),  in the range $3\lesssim q\lesssim100$~\cite{Zappa:2019ntl,Kyutoku:2021icp}.
Hence, ground-based GW interferometers will observe not only their inspiral, but also the last stages of the coalescence, including the merger. 
These binary systems are also promising engines of gamma-ray bursts~\cite{Narayan:1992iy} and kilonovae~\cite{Metzger:2010sy}, which are expected electromagnetic counterparts to future GW events.

The merger process can be quantitatively studied only by means of numerical relativity (NR) simulations~\cite{Taniguchi:2007xm,Kyutoku:2010zd,Kyutoku:2011vz,Shibata:2011jka,Foucart:2013psa,Kyutoku:2015gda,Hinderer:2016eia,Foucart:2018lhe,Chakravarti:2018uyi,foucart_francois_2020_4139881,duez_matthew_2020_4139890,Hayashi:2020zmn}. The tidal disruption of the neutron star (NS) companion plays a key role in the merger dynamics and for determining the GW morphology~\cite{Kyutoku:2011vz,Shibata:2011jka}. The phenomenology observed in simulations indicates the existence of three main the scenarios: in the first, the NS is tidally disrupted before merging with the BH; in the second the NS plunges directly into the BH without getting tidally disrupted; in the third the BH's tidal field induces unstable mass transfer from the NS during mass shedding~\cite{Foucart_2020}. Kyutoku \etal~tentatively classified these cases as mergers of type I, II and III respectively, and highlighted their different imprint in the ringdown part of the GW spectra~\cite{Kyutoku:2011vz,Pannarale:2015jia}. 
Direct plunges produce gravitational waveforms that are similar to BBH ones. 
The fate of the NS has the largest impact  on the morphology of the GW signal. In particular, mass
shedding or tidal disruption causes partial supression of
quasi-normal-modes (QNMs), producing waveforms that deviate 
from BBH ones.

Whether the NS gets tidally disrupted or plunges directly into the BH depends, in the Kerr test-mass approximation, on the mass ratio of the binary, and on the location of the BH's last stable orbit (LSO)\cite{Kawaguchi:2015bwa}. In this case, the ratio between the tidal disruption radius ($r_{\rm{TD}}$) and the LSO radius ($r_{\rm{LSO}}$) scales approximately as $\xi = r_{\rm{TD}}/r_{\rm{LSO}} \propto C^{-1}q^{-2/3}f(\abh)^{-1}$, where $C=G\Mns/(c^2R_{\rm NS})$ is the compactness of the NS of radius $R_{\rm NS}$ and $f(\abh)$ encodes the well-known behaviour of the Kerr LSO as a function of the BH dimensionless and mass-rescaled spin, $\abh$~\cite{Bardeen:1972fi}. Hence, significant mass shedding and tidal disruption ($\xi>1$) is expected for mass ratios approaching one and large NS compactness, while direct plunges ($\xi<1$) will occur in the opposite cases~\cite{Hayashi:2020zmn}. 

Large initial BH spins aligned (anti-aligned) to the orbital angular momentum as well as 
small (large) NS compactnesses favour (disfavour) mass shedding and tidal disruption 
\cite{Duez:2009yy,Etienne:2010ui,Kyutoku:2011vz}. Mass shedding and tidal disruption
lead to the formation of an accretion disk surrounding the remnant BH 
\cite{Pannarale:2012ux,Foucart:2012nc,Kawaguchi:2015bwa,Kruger:2020gig}. 
A fraction of the matter is expected to be ejected on dynamical timescales from the remnant disk, 
undergoing $r$-process nucleosynthesis and thus producing a kilonova 
signal~\cite{Korobkin:2012uy,Rosswog:2013kqa,Hayashi:2020zmn,Fernandez:2016sbf}. 
The remnant BH properties have been shown to depend sensitively on the 
binary properties and to be related to disk formation~\cite{Pannarale:2013jua,Zappa:2019ntl}.

Waveform templates are a necessary ingredient for modelled searches and analyses of GW data. 
\citet{Lackey:2011vz,Lackey:2013axa} designed inspiral-merger-ringdown BHNS waveforms by calibrating the phenomenological BBH model \Phenom{C}~\cite{Santamaria:2010yb} with data from BHNS NR waveforms. 
Following a similar approach,~\citet{Pannarale:2013uoa} developed a phenomenological waveform model for non-spinning binaries that corrects the GW amplitude to account for the three possible merger dynamics described above. The amplitude correction is based on an analytical model of the merger remnant by ~\citet{Pannarale:2012ux}. This model estimates the final BH's mass and spin using the expressions for the LSO of a test-mass orbiting a Kerr BH and NR fits of the remnant disk baryon mass. It was later extended to model the $(\ell,m)=(2,2)$ mode of BHNS with nonprecessing spins using a larger number of NR simulations. ~\citet{Pannarale:2015jia,Pannarale:2015jka} further classify the mergers as disruptive, nondisruptive, and mildly disruptive with and without a disk. 
More recently, ~\citet{Thompson:2020nei} applied the same approach of the above mentioned Pannarale \etal~ to develop the \Phenom{NSBH} approximant that employs a NR-informed closed-form expression for the GW phase contributions due to tidal interactions~\cite{Dietrich:2017aum,Dietrich:2019kaq}. 
The same approach was used to build \SEOB{$\_$NSBH} by~\citet{Matas:2020wab}, in which the \Phenom{C} baseline is replaced with an effective-one-body (EOB) model 
~\cite{Bohe:2016gbl,Dietrich:2019kaq}. In contrast to \Phenom{NSBH}, \SEOB{$\_$NSBH} makes use of additional NR data available for their fitting of amplitude corrections, as well as the 
remnant BH fits of ~\citet{Zappa:2019ntl}. ~\citet{Steinhoff:2021dsn} developed another EOB model as well, that additionally accounts for $f$-mode excitations of the NS through an effective Love number.

In this paper, we propose a new waveform model for waveforms from quasi-circular BHNS systems that we incorporate within the EOB model \TEOB{}, 
previously developed for BBH~\cite{Damour:2014sva,Nagar:2015xqa,Nagar:2018zoe,Nagar:2019wds,Nagar:2020pcj} and binary neutron stars (BNS) 
mergers~\cite{Bernuzzi:2014owa,Akcay:2018yyh}. Our model differs from the above BHNS waveform approximants in the use of NR data. 
Specifically, \TEOB{} does not make use of remnant disk fits from NR, but relies on 
(i) a NR-informed remnant BH model that updates the one from~\citet{Zappa:2019ntl}, 
(ii) next-to-quasicircular corrections (NQC) to the waveform that are specifically designed using NR BHNS simulation data, and 
(iii) a ringdown model based on \TEOB{}' ringdown model for BBH~\cite{Damour:2014yha}. The latter is suitably ``deformed'' to include the BHNS ringdown 
in cases of significant tidal disruption, using a set of pseudo quasi-normal-modes~\cite{Taracchini:2012ig} extracted from the numerical simulations.
Tidal effects are included following the gravitational-self-force resummation prescription for the tidal radial potential of \TEOB{}~\cite{Bernuzzi:2014owa,Akcay:2018yyh}.
Higher modes beyond the dominant quadrupole, $(\ell,m) = (2,2)$, are included when constructing the non-precessing model baseline. 
These are relevant for the description of unequal-mass binaries (as expected for BHNS), we additionally include a straightforward prescription to account for precessing binaries using 
the recipe of Ref.~\cite{Gamba:2021ydi}. We present here the first BHNS waveform model that incorporates higher modes (HMs) and includes spin-precession.

The plan of the paper is as follows. Sec.~\ref{sec:model} describes the new design choices and the implementation
of our model with the available NR simulation data, showing how \TEOB{} for BHNS robustly provides waveforms for a large parameter space spanning mass ratio $q\in[1,20]$ and BH spins as high as $\abh=0.99$.
The waveforms are validated in Sec.~\ref{sec:validation} against NR data as well as other available models. 
Good agreement with NR data is found, with a similar performance to other available models, but significant phasing improvements especially 
in the merger-ringdown of challenging configurations. Additionally, a phasing agreement at a similar level holds even against the only available 
precessing configuration~\cite{chernoglazov_alexander_2020_4139871,Foucart:2020xkt}, a comparison which is performed here for the first time against our model. 
Sec.~\ref{sec:inference} demonstrates the application of \TEOB{} to both artificial and real data, 
including that from the recent BHNS observations~\cite{Abbott_2021}. Using this model with (2,2) mode in 
the analysis of simulated signals gives results very close to the injected values, within statistical 
errors, even for signals with low signal-to-noise-ratio (SNR). This confirms the
self-consistency of the model. For real events, results of the inference are also consistent with the 
ones computed by the LIGO-Virgo-Kagra Collaboration (LVK). We do not find decisive evidence on the 
binary's origin for the analysis on GW190814 and GW200105.
However, we are able to significantly constrain the parameters' values by employing a BHNS model with HMs. 
Similarly for GW200115, we find no evidence of tidal effects.
Finally, in Sec.~\ref{sec:conclu} we discuss the improvements that new NR data would bring to the model's accuracy and suggest the region of the BHNS binary parameter space that should be numerically explored to this aim.\\
  
\paragraph*{Notation.} In this work we use
$M$ for the binary mass, $q=m_1/m_2\geq1$ for the mass ratio, $\nu = q/(1+q)^2$ for the symmetric mass ratio,
$\Mbh,\,\abh$ for the mass and dimensionless spin of the initial BH in the binary system, 
similarly $\Mns,\,a_{\rm{NS}}$ for the NS,
$\Mfbh,\,\afbh$ for the mass and dimensionless spin of the remnant BH. 
The quadrupolar tidal polarizability parameter of the NS is 
$\Lambda = \frac{2}{3}\frac{k_2}{C^5} $, where $k_2$ corresponds to the 
gravito-electric Love number $k_{\ell}$ with $\ell =2$ and $C=M_{\rm{NS}}/R_{\rm{NS}}$ is the compactness of the star. 
The tidal coupling constant parametrizing
 the leading order tidal interactions is 
$\kt2 = \frac{3}{16}\tilde{\Lambda}$ for BHNS, where $\tilde{\Lambda}$ is the reduced tidal deformability defined as~\cite{Lackey:2014fwa},
\be
\tilde{\Lambda} = \frac{8}{13}[1+7\nu-31\nu^2 - \sqrt{1-4\nu}(1+9\nu-11\nu^2)]\Lambda \, .
\ee
The Regge-Wheeler-Zerilli normalized multipolar waveforms are $\Psi_{\ell m}=h_{\ell m}/\sqrt{(\ell +2)(\ell +1)\ell (\ell -1)}$, with the strain multipoles
\be\label{eq:strainh}
h_+ - ih_{\times}=\frac{1}{R}\displaystyle\sum_{\ell=2} ^{\infty}\displaystyle\sum_{m=-\ell} ^{\ell}h_{\ell m}\, _{-2}Y_{\ell m},
\ee
where $R$ is the luminosity distance and $_{-2}Y_{\ell m}$ are the
$s=-2$ spin-weighted spherical harmonics. We 
plot waveforms in terms of the retarded time $u=t-r_*$ typically shifted by the merger time $t_{\rm{mrg}}$. 
The latter is conventionally defined as the time corresponding to the peak amplitude 
of the $(2,\,2)$ mode.

We employ geometric units $c=G=1$ and solar masses $\Msun$, unless explicitly indicated, and use $\log$ to indicate the natural logarithm.

\section{\TEOB{} model}
\label{sec:model}

\TEOB{} is a waveform model based on the EOB formalism~\cite{Buonanno:1998gg,Buonanno:2000ef,Damour:2000we,Damour:2001tu,Damour:2014sva,Damour:2015isa}. 
This approximant produces gravitational waveforms from spinning coalescing compact binaries including higher modes 
and tidal effects~\cite{Nagar:2015xqa,Nagar:2018zoe,Akcay:2018yyh,Nagar:2019wds,Nagar:2020pcj,Riemenschneider:2021ppj}. 
Fast waveform generation is obtained using the post-adiabatic approximation to the EOB Hamiltonian dynamics~\cite{Nagar:2018gnk}. 

The post-merger part of the waveform is characterized by the properties of the final BH, the QNMs driving the relaxation towards equilibrium, and the 
time evolution of modes amplitudes and phases. Due to the different physics close to merger, where finite-size effects start to become relevant, a BHNS 
ringdown differs with respect to the one resulting from a BBH coalescence. The implementation of BHNS coalescences into \TEOB{} thus consists of two main 
parts: the remnant BH characterization and the ringdown model. The idea behind the BHNS modelling is to employ NR fitting formulas describing the deviation 
of a given quantity compared to the BBH case. 

\subsection{Remnant BH model}
\label{sec:model:remnant}

Accurate models for the mass $\Mfbh$ and dimensionless spin
$\afbh$ of the remnant BH are required to construct an EOB model of
the BHNS merger gravitational waveforms. 
As further discussed in Sec.~\ref{sec:model:ringdown}, the remnant's mass and
spin enter the computation of QNMs and other ringdown 
properties of the GW in the post merger phase. In addition, the remnant mass 
is also employed to distinguish among different NS tidal disruption cases.

Fitting formulas for both $\Mfbh$ and $\afbh$ have been developed 
in~\cite{Zappa:2019ntl} using the data from 86 
NR simulations of BHNS mergers published
in~\cite{Kyutoku:2010zd,Kyutoku:2011vz,Kyutoku:2015gda}.  
Those models consist of maps of the type
\be\label{eq:Fmap}
F:\left(\nu,\,a_\text{BH},\,\Lambda\right)\rightarrow\left(\Mfbh,\,\afbh\right)\,,
\ee
which allow to calculate the properties of the remnant BH 
from the binary symmetric mass-ratio $\nu$, the initial BH
spin $\abh$ and the NS' tidal polarizability $\Lambda$. These formulas reduce by construction to the BBH case in the $\Lambda = 0$ limit,
and to the test-mass case in the $\nu \to 0$ limit.

In this paper, we update the models of
Ref.~\cite{Zappa:2019ntl} by including 19 additional simulations of 
non-spinning  BHNS presented in~\cite{Hayashi:2020zmn}. These new
simulations further explore the parameter region of low mass-ratio
binaries ($q\lesssim 3$) for which the NS tidal disruption is more
likely to occur. Clearly, this addition makes the remnant formulas
more accurate in this regime. Also, the fitting formula for
$\Mfbh$ employed here is different from that of
Ref.~\cite{Zappa:2019ntl}; the reason for this choice
is to use a simpler and smoother function in the region of large
$\Lambda$ and small $\nu$, where no NR data are available.
The remnant's fitting model (Eq.~\eqref{eq:Ffit1} for $\afbh$ and Eq.~\eqref{eq:Ffit2} for $\Mfbh$ with $\lambda=\Lambda$)  and the
updated coefficients are reported in
Appendix~\ref{app:fitmodel}. 

\begin{figure*}[t]
  \centering
  \includegraphics[width=0.9\textwidth]{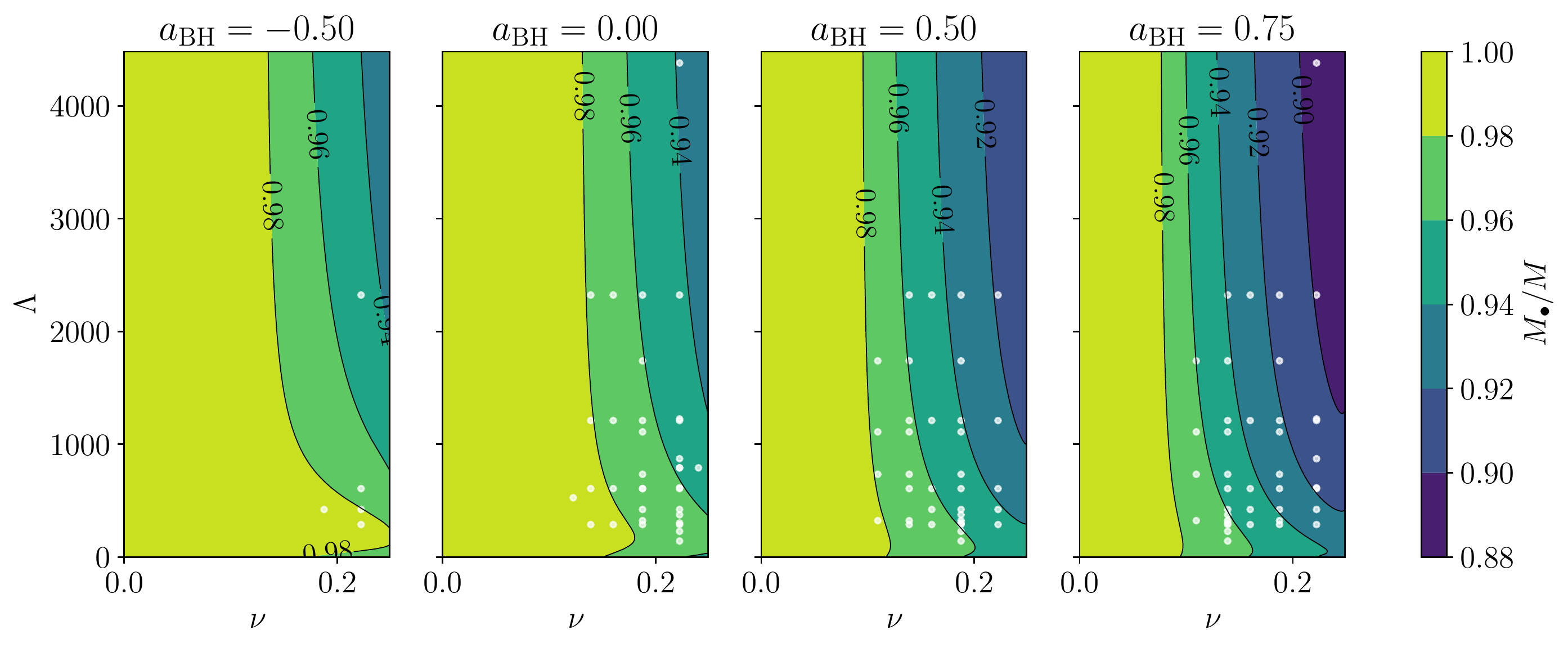}
  \caption{BH remnant mass fit $\Mfbh{}/M(\nu,\Lambda,\abh)$ for
    representative values of the BH spin $\abh$.
    NR data points are shown as white dots.
     }
  \label{fig:Mfbhfit}
\end{figure*}

Representative plots of the $M$ rescaled remnant mass $\Mfbh{}/M$ are shown in Fig.~\ref{fig:Mfbhfit} for different values of spin $\abh$. 
The lowest values of $\Mfbh{}/M$ are attained in the high $\Lambda$ - high $\nu$ regime, where tidal effects play a significant role in the dynamics. Consequently, the NS mass does not contribute much to the final BH mass due to disk formation through tidal disruption. 
On the other hand, a clear mass peak is present  at low values of $\Lambda$, and all values of $\nu$. Since tidal interactions are negligible in this region, there is no disk formation and the NS mass gets swallowed by the BH. 
With increasing spin $\abh$, the parameter $\xi$ that controls the onset of tidal disruption increases due to the repulsive spin-orbit interaction. 
Indeed, lower values of $\Mfbh{}/M$ are attained when the spin is large, and the mass peak in the low $\Lambda$ region is suppressed.

\subsection{Ringdown model}
\label{sec:model:ringdown}

The ringdown emission in BHNS mergers must account for the NS tidal
disruption that, in certain binary parameters regions, significantly
suppresses QNM excitation. In these regions, the BBH ringdown
representation is no longer accurate.
The BHNS ringdown model is constructed as a deformation of the EOB
ringdown for BBHs proposed in
Ref.~\cite{Damour:2014yha}. The ringdown of a GW mode is analytically
modelled by the multiplicative ansatz
\be\label{eq:hrd}
\bar{h}_{\lm}(\tau)= e^{\sigma_1 \tau + i \phi _{0}}h(\tau)\equiv A_{\bar{h}(\tau)}e^{i\phi _{\bar{h}}(\tau)},
\ee
where $\sigma_1=\alpha_{\lm 1}+i\omega_{\lm 1}$ is the dimensionless complex
frequency of the fundamental QNM ($n=1$) with the
inverse damping time $\alpha_{1}$ and frequency $\omega_{1}$ [hereafter the ($\ell$, $m$) indices are
  often omitted], 
and $\tau = (t-t_0)/\Mfbh$ is
a dimensionless time parameter. The quantities $A_{\bar{h}(\tau)}$
and $\phi_{\bar{h}}(\tau)$ are written as
\begin{subequations}
\label{eq:amppha}
\begin{align}
A_{\bar{h}(\tau)}&=c^A_1\tanh(c^A_2\tau + c^A_3)+c^A_4,\\
\phi _{\bar{h}}(\tau)&=-c^{\phi}_1\ln \left( \frac{1+c^{\phi}_3e^{-c^{\phi}_2\tau}+c^{\phi}_4e^{-2c^{\phi}_2\tau}}{1+c^{\phi}_3+c^{\phi}_4} \right).
\end{align}
\end{subequations}
A set of physical constraints are imposed to the parameters
$c^{A,\phi}_i$~\cite{Damour:2014yha}, 
\begin{subequations}
\label{eq:postpeakcoef}
\begin{align}
c^A_2&=\frac{1}{2}\alpha_{21},\\
c^A_4&=\hat{A}^{\rm{peak}}-c^A_1\tanh(c^A_3),\\
c^A_1&=\hat{A}^{\rm{peak}}\alpha _1\frac{\cosh ^2(c^A_3)}{c^A_2},\\
c^{\phi}_1&=\frac{1+c^{\phi}_3+c^{\phi}_4}{c^{\phi}_2(c^{\phi}_3+2c^{\phi}_4)}(\omega _1-\Mbh\omega^{\rm{peak}}),\\
c^{\phi}_2&=\alpha _{21},
\end{align}
\end{subequations}
with $\alpha_{21}\equiv \alpha_2 - \alpha_1$, where $\alpha_2$
corresponds to the inverse damping time of the first overtone ($n=2$).
Therefore, three free parameters remain: $c^A_3$, $c^{\phi}_3$, $c^{\phi}_4$. 

The ringdown model of Eq.~\eqref{eq:hrd} thus requires (for each multipole)
the QNM frequencies $(\omega_1,\alpha_1,\alpha_2)$, the peak (or maximum) values
of the amplitude $\hat{A}^{\rm{peak}}$ and frequency $\omega ^{\rm{peak}}$ entering
Eq.~\eqref{eq:postpeakcoef} and the parameters $(c^A_3, c^{\phi}_3, c^{\phi}_4)$. 
The latter are fit to BBH NR data in~\cite{Nagar:2020pcj} for both the $(2,2)$ and higher modes, while the set
\be\label{eq:rng:fit:pars}
(\omega_1,\, \alpha_1,\, \alpha_2,\, \hat{A}^{\rm{peak}},\, \omega^{\rm{peak}}) 
\ee
is modelled using the BHNS NR data available to us, see
App.~\ref{app:NRdata}. 
The quantities in \eqref{eq:rng:fit:pars} are
modelled similarly to the BH remnant properties in \S\ref{sec:model:remnant}.
The QNM quantities $(\omega_1,\, \alpha_1,\, \alpha_2)$ are fit using
Eq.~\eqref{eq:Ffit1} with $\lambda=\Lambda$ and correspond to pseudo-QNMs~\cite{Taracchini:2012ig}.
The quantities $(\hat{A}^{\rm{peak}}$, $\omega^{\rm{peak}})$
are modelled with Eq.~\eqref{eq:Ffit2} and employ the tidal coupling constant $\lambda=\kappa^{\rm{T}}_2$ instead of $\Lambda$.
The underlying BBH values for the above parameters are those of
Ref.~\cite{Nagar:2020pcj}, to which the model exactly reduces for
$\lambda=0$ and $\nu\to0$.  
Table~\ref{tab:fpeaksco} summarizes the results for the all the
fitting parameters for the ($\ell=2$,\,$m=2$) mode.
Our model also includes HMs to account for their significant contribution to the waveform morphology for increasing mass ratio. Due to the lack of enough simulations including HMs, we apply the corrections obtained for (2,2) mode to the corresponding BBH fits for subdominant modes.
The current available modes for our BHNS model are: (2,1), (2,2), (3,2), (3,3), (4,4) and (5,5).

\begin{figure*}[t]
  \centering
  \includegraphics[width=0.9\textwidth]{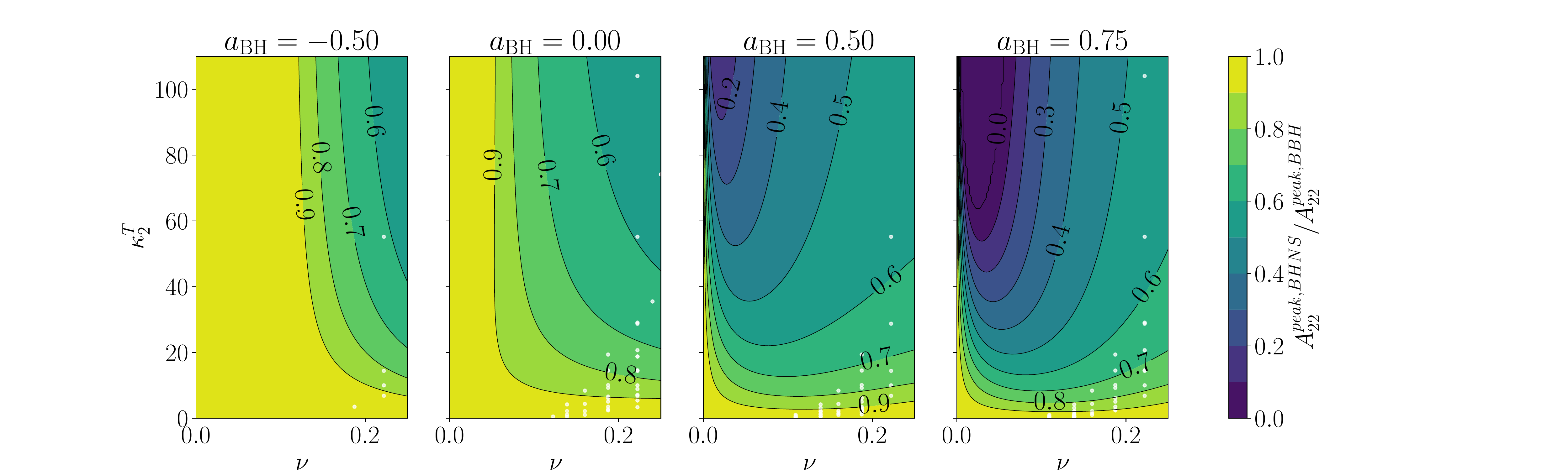}
  \includegraphics[width=0.9\textwidth]{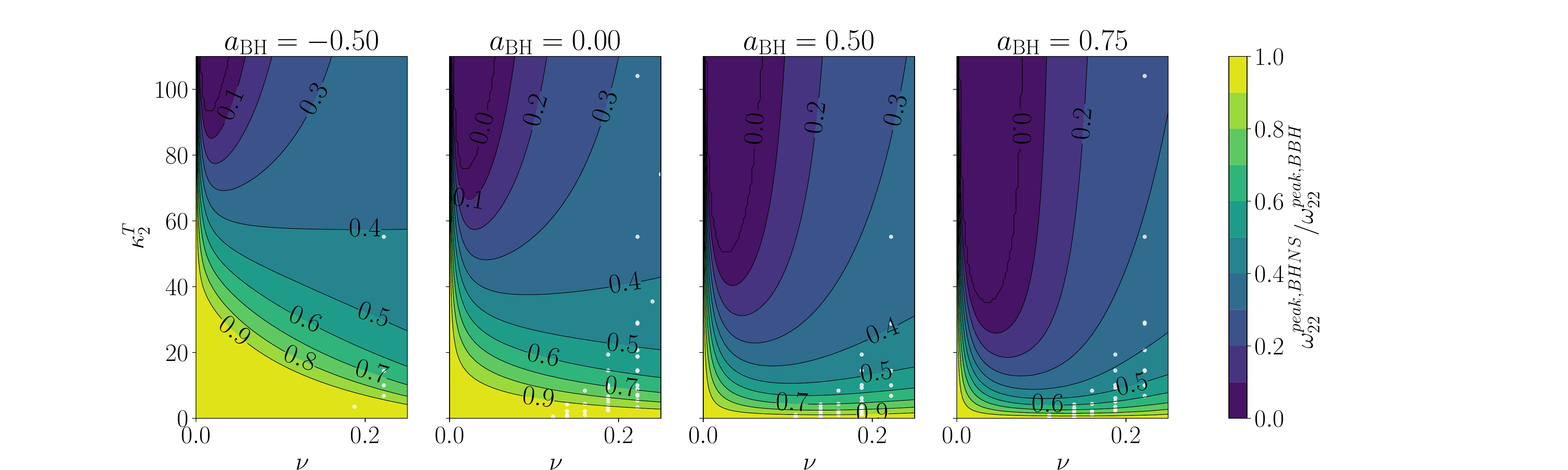}
  \includegraphics[width=0.9\textwidth]{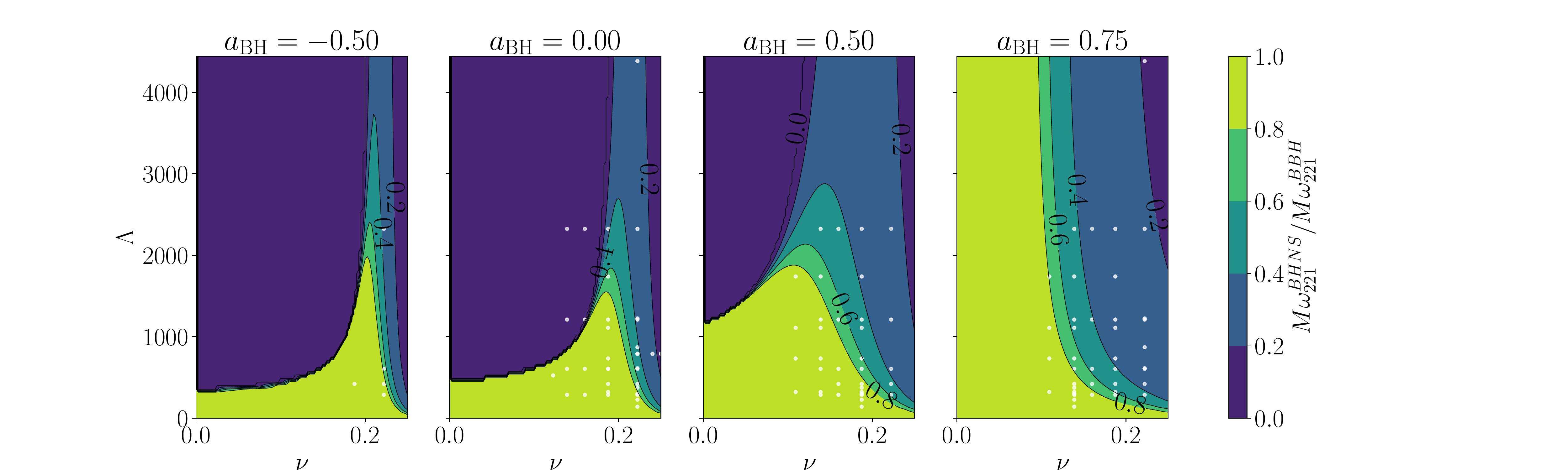}
  \caption{Two-dimensional plots of the fits for the BHNS ringdown quantities, relative to the BBH values: $A^{\rm{BHNS}}_{22}/A^{\rm{BBH}}_{22}$(top), $\omega^{\rm{BHNS}}_{22}/\omega^{\rm{BBH}}_{22}$ (middle) and $M \omega^{\rm{BHNS}}_{221}/M \omega^{\rm{BBH}}_{221}$ (bottom). These are plotted with representative values of the progenitor BH spin $\abh$ (see text). The NR data employed in the fits are represented with white dots.}
  \label{fig:2Dfits}
\end{figure*}

Figure~\ref{fig:2Dfits} shows $A^{\rm{peak,BHNS}}_{22}/A^{\rm{peak,BBH}}_{22}$ (top), $\omega^{\rm{peak,BHNS}}_{22}/\omega^{\rm{peak,BBH}}_{22}$ (middle) and $M\omega^{\rm{BHNS}}_{221}/M\omega^{\rm{BBH}}_{221}$ (bottom) for representative values of $\abh$ (in the latter, the indices correspond to $\ell$, $m$ and the $n^{th}$ overtone). 
For a progenitor BH with $\abh=0$, tidal disruption significantly suppresses the peak amplitude for $\nu>0.18$, whereas for $\nu<0.16$ and low $\kappa^{\rm{T}}_2$ values, it is well represented by the BBH model. 
When the spin increases, we observe a stronger suppression for a wider range of mass ratios. 
This is a consequence of the repulsive spin-orbit terms in the dynamics, that effectively reduce the radius of the black hole's innermost stable circular orbit (ISCO) 
for increasing spin. 
Therefore, the onset of tidal disruption, $\xi$, occurs before the NS gets swallowed by the BH. 
For this reason, BHNS coalescences with a high spinning BH cannot be faithfully described by a BBH model. 
For the peak frequency $\omega^{\rm{peak}}_{22}$, we observe a similar behaviour with increasing $\abh$.
Only for very low values of $\kappa^{\rm{T}}_2 \lesssim 10$ we obtain peak frequencies that mimic those of the BBH case. 
The QNM frequency $M\omega_{221}$ shows a different behaviour, which however we believe is due to the lack of data in the low $\nu$ region. 
Indeed, for high $\nu$ values, we obtain a behaviour consistent with the other quantities, with increasing $\abh$ suppressing the frequency due to tidal disruption.

\begin{figure*}[t]
  \centering
  \includegraphics[width=0.9\textwidth]{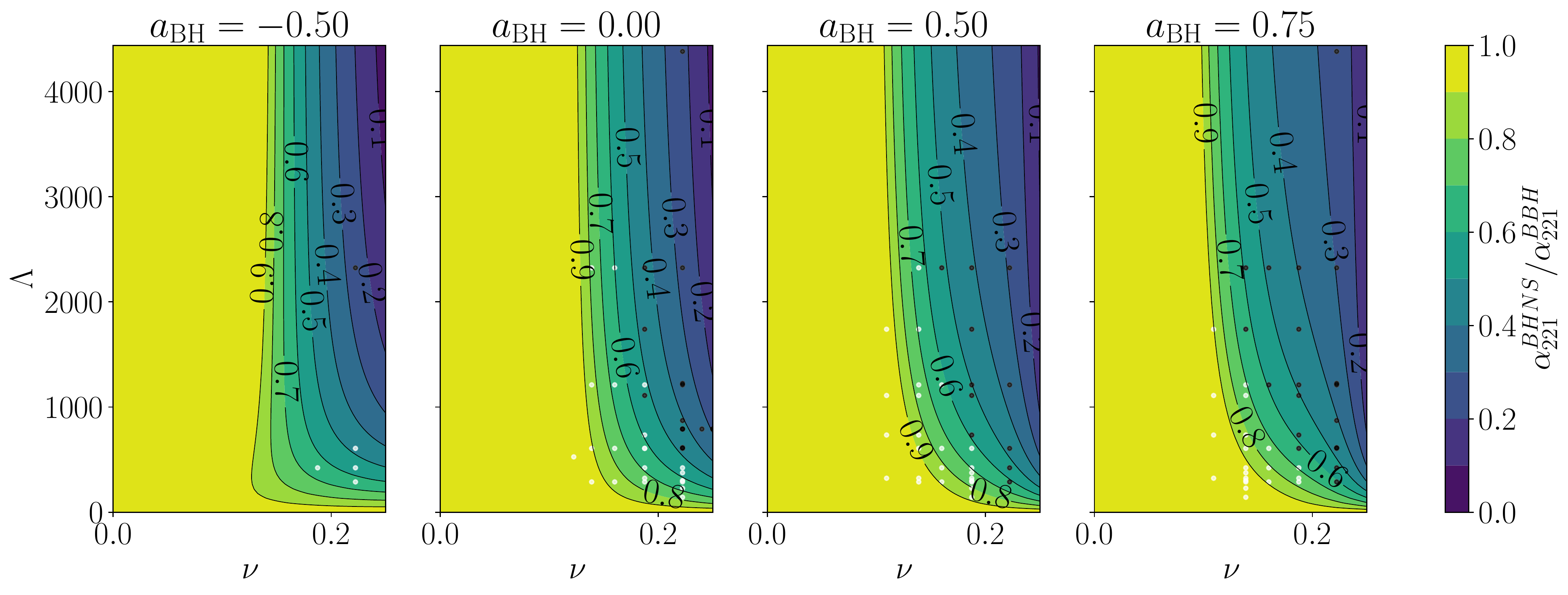}
  \caption{Two-dimensional plots of the fits for the inverse damping time $\alpha_{221}(\nu,\Lambda,\abh)$. The white dots represent the NR data, the ones marked in black are the tidally disrupted cases (see text).}
  \label{fig:2Dfits_idt}
\end{figure*}

Fig.~\ref{fig:2Dfits_idt} shows the fits for the inverse damping time. 
To recognize the tidally disrupted NS (black dots) from the others (white dots), a criterion based on the contours of this figure is discussed in Sec.~\ref{sec:model:IMR}. 
The tidal disruption cases are identified by the lack of excited QNM in their frequency after merger, in contrast to BBH. 
The BHNS QNM fits developed ($\omega_1$, $\alpha_1$, $\alpha_2$) are used only for BHNS binaries where the NS 
is tidally disrupted, whereas the original QNM fits from~\cite{Nagar:2020pcj} are employed for the rest of the binaries (see also Sec.~\ref{sec:model:IMR}). 

\subsection{Inspiral-merger-ringdown waveforms}
\label{sec:model:IMR}

\TEOB{} inspiral-merger-ringdown waveforms (IMR) for BHNS are constructed
exactly as BBH waveforms, but using: 
(i) the GSF3 model of~\cite{Bernuzzi:2014owa} for gravitoelectric 
 tidal effects and the Post Newtonian (PN) model of~\cite{Akcay:2018yyh} for gravitomagnetic tidal effects,
(ii) the remnant model described in Sec.~\ref{sec:model:remnant} and
(iii) the ringdown model described in Sec.~\ref{sec:model:ringdown}. 
In addition, NQC corrections to the waveform differ from the BBH case, and are computed for each multipole. 
The NQC extraction points from~\cite{Nagar:2020pcj} are employed
, with additional BHNS fits for
\be
(A^{\rm{NQC}}_{22},\, \dot{A}^{\rm{NQC}}_{22},\, \omega^{\rm{NQC}}_{22},\, \dot{\omega}^{\rm{NQC}}_{22}) 
\ee
for the cases that deviate sufficiently from the BBH case. The BHNS NQC fits were modelled
in the same manner as the ringdown fits of the previous section. The model employed for the quantities ($\dot{A}^{\rm{NQC}}_{22}$, $\omega^{\rm{NQC}}_{22}$) is Eq.~\eqref{eq:Ffit1}, while for ($A^{\rm{NQC}}_{22}$, $\dot{\omega}^{\rm{NQC}}_{22}$) Eq.~\eqref{eq:Ffit2} is used, with $\lambda=\kappa^{\rm{T}}_2$ in both cases. More details on the fits can be found in Appendix~\ref{app:fitmodel}.
Tidal disruption in BHNS mergers can be roughly classified in three types~\cite{Kyutoku:2011vz}:
(I) the NS is tidally disrupted far away from the ISCO, and the ringdown is strongly suppressed;
(II) the NS is not tidally disrupted, and the ringdown is very similar to the BBH case;
(III) the NS is tidally disrupted close to the ISCO, and the ringdown is present, but significantly altered by the tidal disruption.
Given the scarcity of NR data required to develop remnant and ringdown models on the entire BHNS parameter space, the \TEOB{} BHNS model currently implements the above classification.

The three cases (I-III) are quantitatively identified in the parameter space using the QNM damping time model of Sec.\ref{sec:model:ringdown}, and are shown in Fig.~\ref{fig:2Dfits_idt}. 
Among all the fits developed for the BHNS model, this parameter is the one that better allows to distinguish the tidal disruption cases from the rest. The QNMs 
damping times allow to capture the impact on the BHNS ringdown morphology induced by the reduced post-merger emission.
Additionally, these fits also show a more physical behaviour throughout the parameter space, e.g. closer to BBHs for low $\nu$, and a growing tidally disrupted 
region with increasing $\nu$ and $\Lambda$. 

\begin{figure*}[t]
  \centering 
  \includegraphics[width=\textwidth]{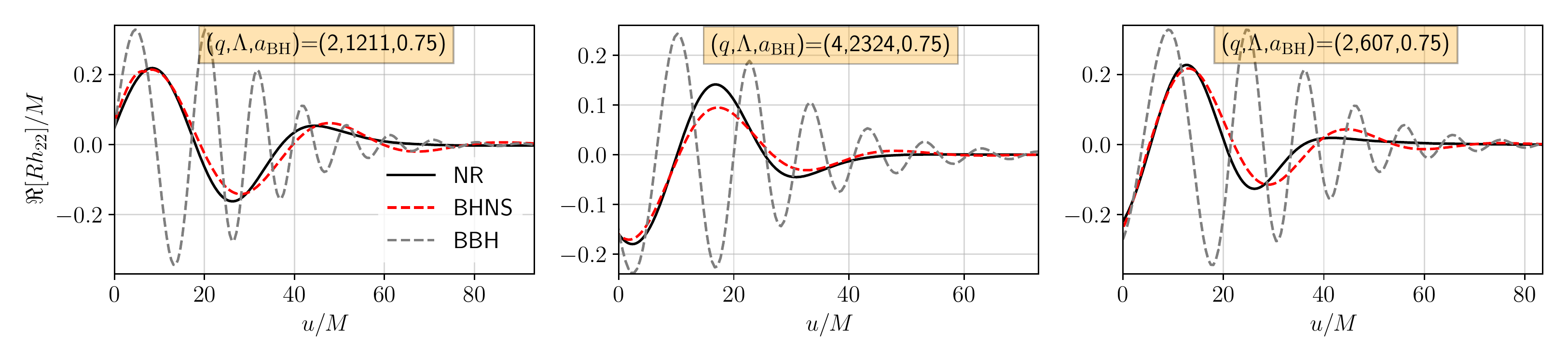}
  \includegraphics[width=\textwidth]{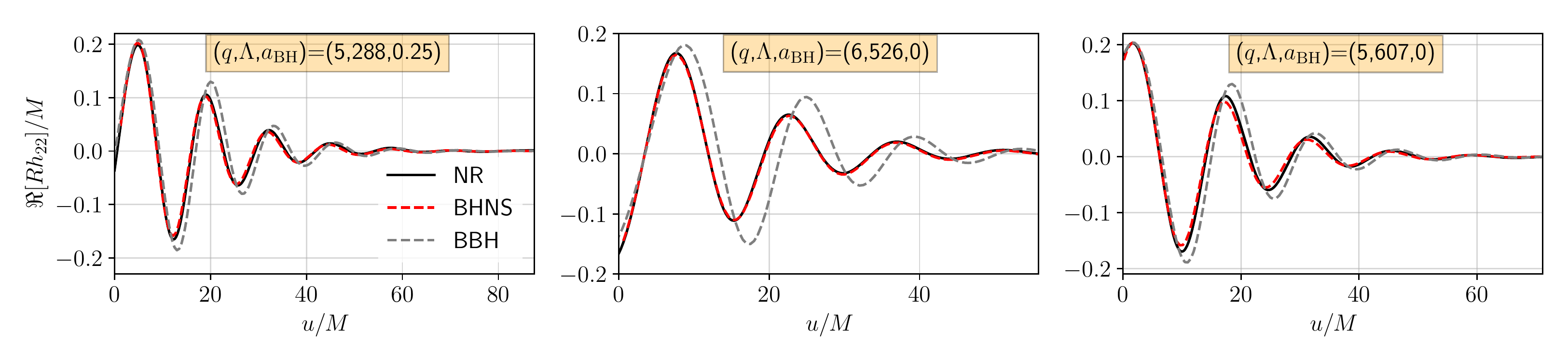}
  \includegraphics[width=\textwidth]{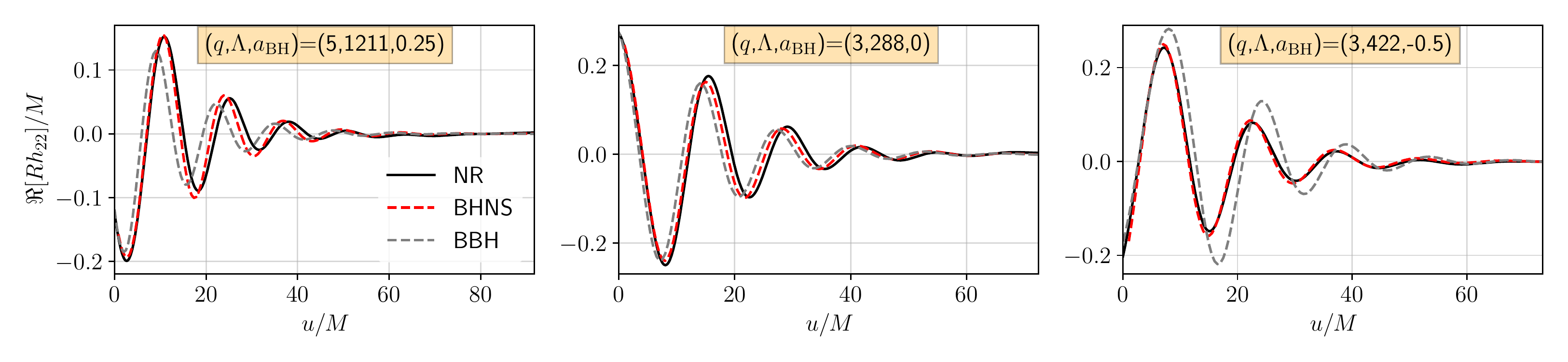}
  \caption{Ringdown alignment at merger for different models with parameters $(q,\Lambda,\abh)$, corresponding to Type I (top), Type II (middle), and Type III (bottom) of our classification (see text). The BBH ringdown model (gray dashed line) is compared directly against our BHNS model (red dashed line), illustrating how the BHNS ringdown is deformed from the BBH one to account for tidal effects.}
  \label{fig:rdfigs}
 \end{figure*}

Type~I binaries are shown as black markers in Fig.~\ref{fig:2Dfits_idt}. The contour around
$\alpha^{\rm{BHNS}}/\alpha^{\rm{BBH}}<0.6$ for nonspinning binaries is adopted as a criterion for identifying Type~I binaries. 
These types of binaries make use of all fits developed for BHNS as described in Sec.~\ref{sec:model:ringdown}.
Type~II binaries have a ringdown that is practically indistinguishable from BBH, and are identified through the contour $\alpha^{\rm{BHNS}}/\alpha^{\rm{BBH}}>0.9$. For these binaries the EOB model simply employs the BBH baseline with the remnant BH model of Sec.~\ref{sec:model:remnant}. 
Type~III binaries are the intermediate cases for which we find sufficient to use a BBH ringdown model, where initial conditions are modified by the
presence of the NS. Hence, the corrections to the peak quantities ($\hat{A}^{\rm{peak}}$, $\omega^{\rm{peak}}$) developed in Sec.~\ref{sec:model:ringdown} are employed for the values and the BHNS NQC parameters described above.
The QNM quantities ($\omega_1$, $\alpha_1$, $\alpha_2$) are instead modelled as in the BBH case~\cite{Nagar:2020pcj}.
Fig.~\ref{fig:rdfigs} shows examples of the different BHNS ringdown 
types against their BBH equivalent. 
Type~I binaries (top) show an almost completely suppressed ringdown.
The QNM fits developed for these cases in Sec.~\ref{sec:model:ringdown} are in good agreement with the NR waveform, especially at the dampened end. 
For Type~II binaries (middle) the BBH fits capture already quite accurately the NR ringdown, confirming the validity of our approach. A slight deformation from the BBH case can be observed, and is due to the different values of the remnant parameters employed.
In Type~III cases (bottom), the BHNS corrections to their respective BBH fits improve the waveform morphology compared to NR. Since these cases use the QNM fits for BBH, they show a small deviation from the BBH ringdown. Appreciable differences can however be noted especially around the merger time, where the impact of our modelling choices is stronger.

\section{Waveform model validation}
\label{sec:validation}

In this section we assess 
the accuracy of our model by comparing it
directly to available NR simulations, as well as to 
other BHNS models, namely \Phenom{NSBH}~\cite{Thompson:2020nei} 
and \SEOB{\_NSBH}~\cite{Matas:2020wab}, 
in addition to \SEOB{NRv4T}~\cite{Steinhoff:2021dsn} 
for relevant cases. 
The comparisons are
performed through waveform phasing and 
unfaithfulness calculations. 

Moreover, the robustness of the model was tested by successfully generating $10^6$ waveforms in a large parameter space. 
Taking into account \TEOB{}'s own regime of validity and expected NS masses, we determine the model's validity to lie within the ranges $q\in[1,20]$, $M_{\rm{NS}}\in[1,3]$, $\Lambda\in[2,5000]$, and $\abh\in[-0.99,0.99]$.

\subsection{NR Phasing}
\label{sec:validation:phasing}

The alignment procedure is carried out as proposed in~\cite{Boyle:2008ge}. For two given
NR and EOB waveforms, we define a function $\Xi^2 (\delta t, \delta \phi)$ of the NR and EOB waveforms phases 
$\phi_{\rm{NR}}$ and $\phi_{\rm{EOB}}$, respectively,
\be\label{eq:ph1}
\Xi^2 (\delta t, \delta \phi)=\int ^{t_f}_{t_i}[\phi_{\rm{NR}}(t)-\phi_{\rm{EOB}}(t+\delta t)+\delta\phi]^2dt,
\ee
where $t_i$ and $t_f$ define the alignment window and time range in which the waveforms are compared. One seeks to minimize this function by finding the optimal values for $\delta t$ and $\delta \phi$. 
The latter are then used to shift the EOB waveform as 
\be\label{eq:ph2}
h_{\rm{EOB}}(t)=A_{\rm{EOB}}(t+\delta t)e^{-i[\phi_{\rm{EOB}}(t+\delta t)+\delta\phi]}
\ee
in order to 
compare it directly to the NR waveform, $h_{\rm{NR}}(t)$.\\

\begin{figure*}[t]
  \centering 
  \includegraphics[width=\textwidth]{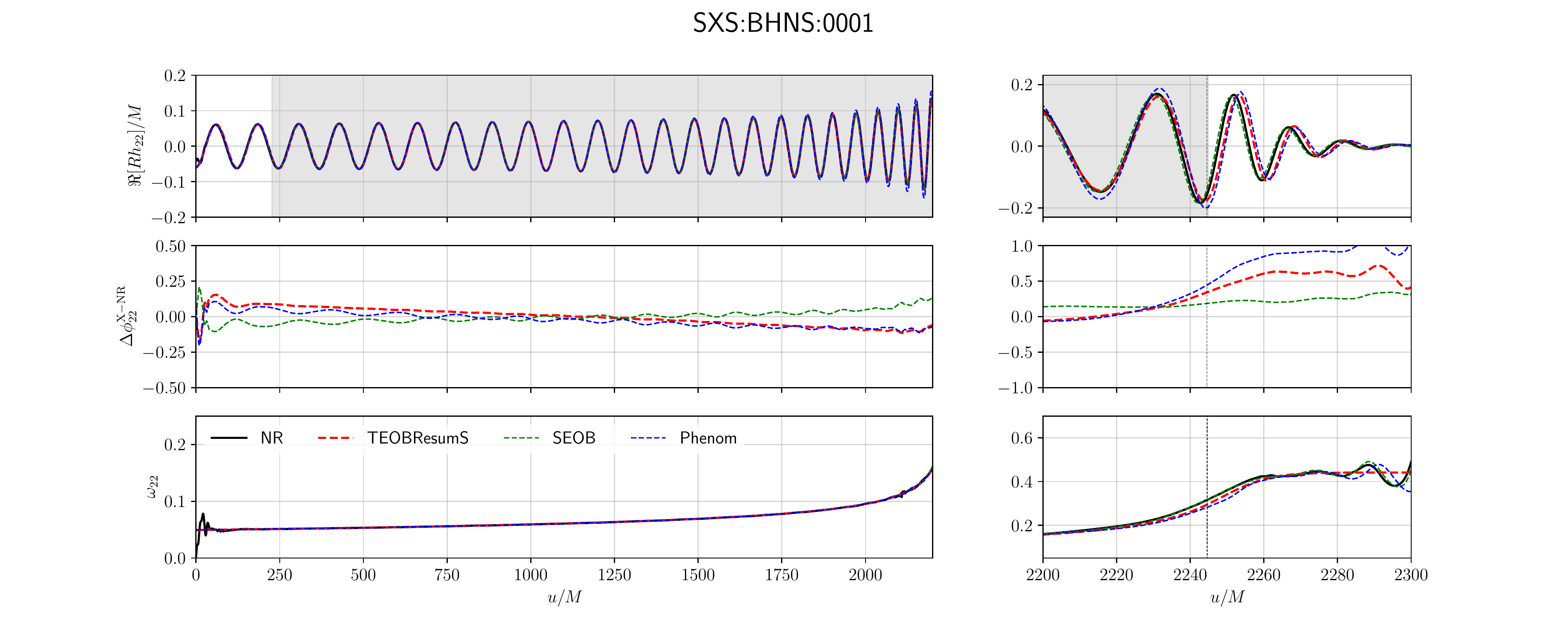}
  \caption{Phasing analysis of the $\ell = m = 2$ mode of SXS:BHNS:0001 against our \TEOB{} model and other available approximants: \SEOB{\_NSBH} and \Phenom{NSBH}. In our classification, this is considered as a Type II binary. Since the NR error of this model is not available, it is not shown. The plot consists on the waveform (top), the phase difference with NR in radians (middle), and the frequency (bottom). The black dashed line indicates the point of merger. The alignment is done by minimizing the phase difference in the gray shaded time window.}
\label{fig:phasing_nospin01}
\end{figure*}

\begin{figure*}[t]
  \centering 
  \includegraphics[width=\textwidth]{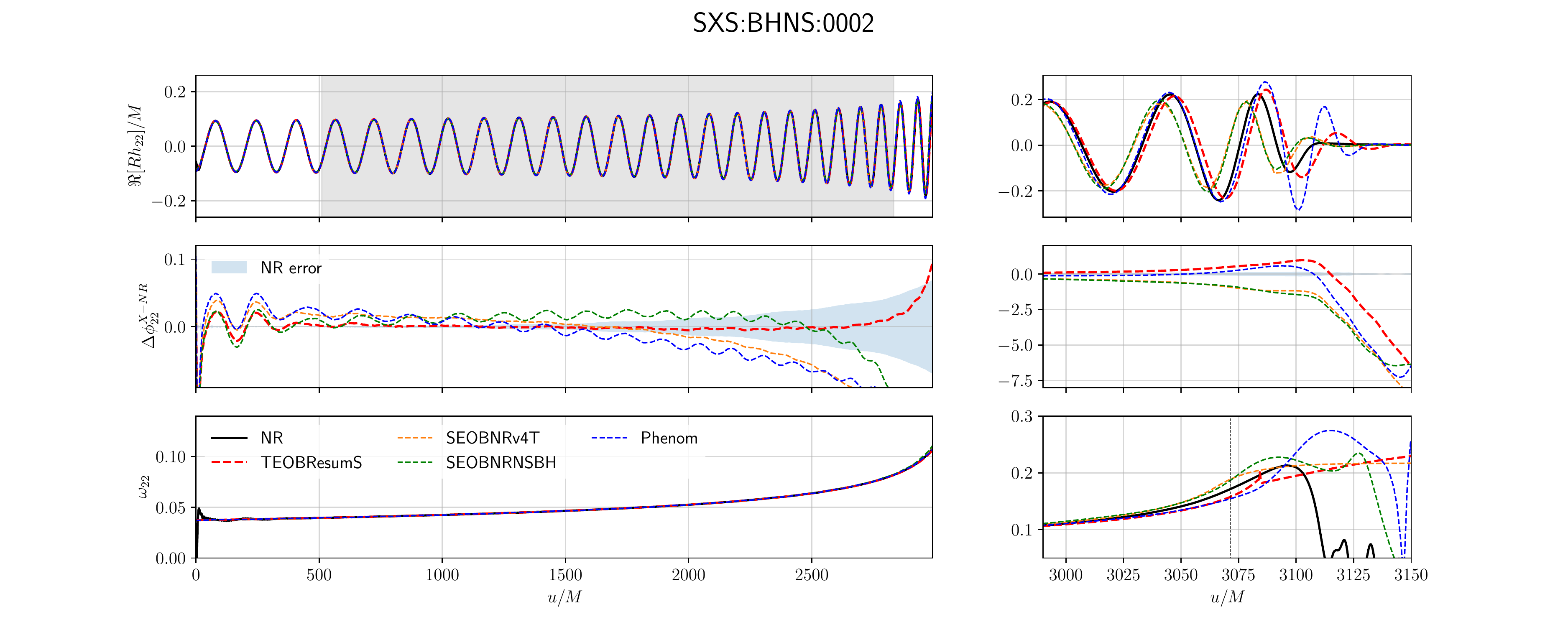}
  \caption{Same as Fig. \ref{fig:phasing_nospin01} but for SXS:BHNS:0002 and including the recent \SEOB{v4T} approximant. In our classification, this is considered as a Type I binary. The NR phase error is shown as a light blue band.
  }
\label{fig:phasing_nospin02}
\end{figure*}

\begin{figure*}[t]
  \centering 
  \includegraphics[width=\textwidth]{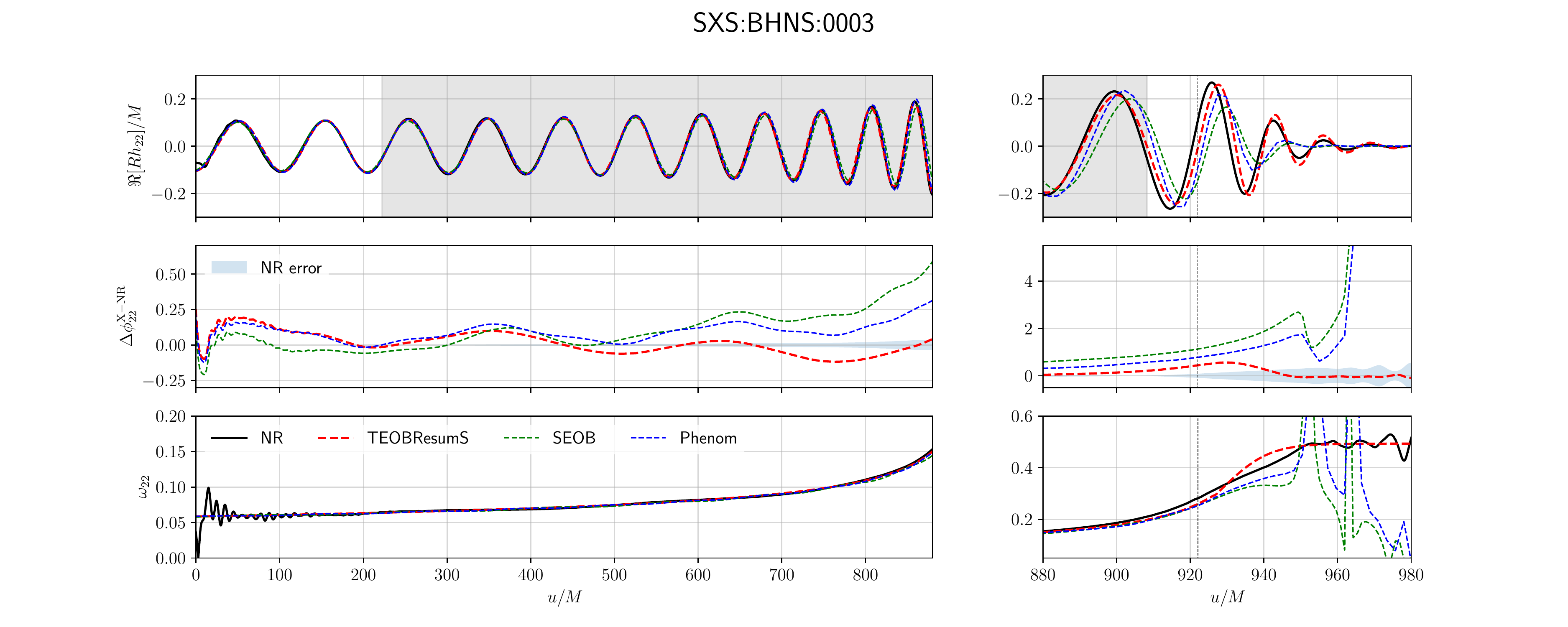}
  \caption{Same as Fig. \ref{fig:phasing_nospin01} but for SXS:BHNS:0003. In our classification, this is considered as a Type III binary. The NR phase error is shown as a light blue band.}
\label{fig:phasing_nospin03}
\end{figure*}

Figures~\ref{fig:phasing_nospin01}, \ref{fig:phasing_nospin02}, and \ref{fig:phasing_nospin03} show the phasing of the model against the NR simulations SXS:BHNS:0001 (Type II),
SXS:BHNS:0002 (Type I), and SXS:BHNS:0003 (Type III) for the dominant (2,2) mode. The gray shaded region marks the time window used for alignment. The alignment was done up to merger for all approximants.
The NR error is shown as a light blue band, in the middle panel of each figure reporting the dephasing error. This is computed 
as the difference
between the phases of the two highest resolutions available.
For the case of SXS:BHNS:0001 only one resolution is available, 
thus the error is not shown in the plot. For 
SXS:BHNS:0001 and SXS:BHNS:0002, the phase difference between NR and 
\TEOB{} is sufficiently small through all inspiral until merger, 
where the dephasing occurs. 
SXS:BHNS:0003 being a much shorter 
waveform shows slight dephasing through the inspiral phase, but an 
overall good agreement with NR even after merger, in contrast to 
the other approximants. Overall, both \SEOB{\_NSBH} and \Phenom{NSBH} show 
a greater dephasing even during the late inspiral. On the other 
hand, the recent tidal approximant \SEOB{v4T} shown in Fig.
\ref{fig:phasing_nospin02} shows a slight dephasing towards merger 
similar to \SEOB{\_NSBH}, but a larger disagreement after merger.\\

Table~\ref{tab:dephase} shows the dephasing for all SXS BHNS models with \TEOB{}. With the exception of SXS:BHNS:0005 and SXS:BHNS:0008, the phase difference at merger stays within less than a radian for all cases. 

\begin{table}[t]
   \centering    
   \caption{Accumulated phase difference between \TEOB{} and NR simulations. Note that the error for SXS:BHNS:0001 is not reported since only one simulation is available.}
   \begin{tabular}{cccc}        
     \hline\hline
     Model & At merger & Total & NR error \\
     \hline
     SXS:BHNS:0001 & -0.339 & -0.640 & - \\
     SXS:BHNS:0002 & -0.530 & -0.406 & 0.118  \\
     SXS:BHNS:0003 & -0.472 & 0.110 & -0.091  \\
     SXS:BHNS:0004 & -0.901 & -1.346 & 0.014 \\
     SXS:BHNS:0005 & -2.224 & -6.179 & -0.063  \\
     SXS:BHNS:0006 & -0.583 & -0.001 & -0.604  \\
     SXS:BHNS:0007 & 0.029 & 1.063 & -0.438  \\
     SXS:BHNS:0008 & -1.402 & -0.376 & 0.046  \\
     SXS:BHNS:0009 & -0.362 & -1.220 & -0.062 \\
     \hline\hline
   \end{tabular}
  \label{tab:dephase}
\end{table}
  
In addition to the above, we also compare NR with \TEOB{} HMs waveforms: (2,1), (3,2), (3,3), (4,4), and (5,5), at $N=3$ extrapolation order.
Examples are shown in Fig.~\ref{fig:phasing_HM} of Appendix~\ref{app:HM}. In general, our model shows good agreement throughout all inspiral for most cases. 
However, given the low resolution of these simulations, one can notice some dephasing before merger and a large NR error towards the ringdown.

\subsection{Faithfulness}
\label{sec:validation:faithfulness}
For a direct comparison with NR, we compute the unfaithfulness defined between EOB and NR waveforms as
\be\label{eq:unfaith}
\bar{\mathcal{F}}\equiv 1-\mathcal{F}=1-\mathop{\rm{max}}_{t_0,\phi_0}\frac{\langle h^{\rm{EOB}},h^{\rm{NR}}\rangle}{\|h^{\rm{EOB}}\| \|h^{\rm{NR}}\|},
\ee
where $t_0$ and $\phi_0$ denote the initial time and phase, and
$\|h\|\equiv \sqrt{\langle h,h\rangle}$. The inner product in Eq.~\eqref{eq:unfaith}
is defined as
\be\label{psd}
\langle h_1,h_2\rangle \equiv 4\Re{ \int \frac{\tilde{h}_1(f)\tilde{h}^*_2(f)}{S_n(f)} }df
\ee
where $S_n(f)$ is the power spectral density (PSD) of the detector and $\tilde{h}(f)$ the Fourier transform of $h(t)$. 
The unfaithfulness, computed both until merger ($f_{\rm{high}}=f_{\rm{mrg}}$) and throughout the coalescence ($f_{\rm{high}}=2048$ Hz) was obtained for each of the (2,2) mode BHNS models available, and is reported in Table~\ref{tab:matching}.

We use the initial frequencies $f_0$ shown in the table and the rest of the binary's parameters for each model are taken from the NR data. The unfaithfulness was obtained using the zero-detuned, high-power Advanced LIGO\cite{aLIGODesign_PSD} noise curve. The results
shown in the table show that the performance of the \TEOB{} BHNS model is comparable to the other approximants especially before merger. 

 \begin{table*}[t]
  \centering    
   \caption{Unfaithfulness of available (2,2) mode BHNS models with long SXS NR waveforms. The results are obtained using the Advanced LIGO zero-detuning, high-power noise curve. Here ``UM'' stands for ``until merger'' and ``FW'' for ``full waveform''.}
   \begin{tabular}{c|c|cc|cc|cc|cc}        
     \hline\hline
     \multirow{2}{*}{Model} & $f_0$ [Hz] & 
     	\multicolumn{2}{c|}{\TEOB{}} & 
     	\multicolumn{2}{c|}{\SEOB{\_NSBH}} & 
     	\multicolumn{2}{c|}{\Phenom{NSBH}} &
     	\multicolumn{2}{c}{\SEOB{NRv4T}}\\
     	
     & &  UM & FW & UM & FW & UM & FW & UM & FW\\
     \hline
     SXS:BHNS:0001 & 169 & 9.8$\times$10$^{-3}$ & 1.0$\times$10$^{-2}$ & 8.4$\times$10$^{-3}$ & 8.4$\times$10$^{-3}$ & 1.0$\times$10$^{-2}$ & 1.1$\times$10$^{-2}$ & 9.9$\times$10$^{-3}$ & 1.3$\times$10$^{-2}$ \\
     
     SXS:BHNS:0002 & 315 & 5.9$\times$10$^{-3}$ & 6.4$\times$10$^{-3}$ & 7.7$\times$10$^{-3}$ & 7.7$\times$10$^{-3}$ & 6.7$\times$10$^{-3}$ & 7.2$\times$10$^{-3}$ & 6.4$\times$10$^{-3}$ & 7.1$\times$10$^{-3}$ \\
     
     SXS:BHNS:0003 & 407 & 7.7$\times$10$^{-3}$ & 7.7$\times$10$^{-3}$ & 9.4$\times$10$^{-3}$ & 9.5$\times$10$^{-3}$ & 1.2$\times$10$^{-2}$ & 1.2$\times$10$^{-2}$ & 2.1$\times$10$^{-2}$ & 2.1$\times$10$^{-2}$ \\
     
     SXS:BHNS:0004 & 447 & 1.3$\times$10$^{-2}$ & 1.3$\times$10$^{-2}$ & 8.4$\times$10$^{-3}$ & 1.2$\times$10$^{-2}$ & 9.9$\times$10$^{-3}$ & 1.6$\times$10$^{-2}$ & 2.4$\times$10$^{-2}$ & 2.4$\times$10$^{-2}$ \\
     
     SXS:BHNS:0005 & 448 & 6.2$\times$10$^{-2}$ & 7.7$\times$10$^{-2}$ & 7.0$\times$10$^{-2}$ & 9.6$\times$10$^{-2}$ & 7.3$\times$10$^{-2}$ & 1.0$\times$10$^{-1}$ & 6.6$\times$10$^{-2}$ & 7.8$\times$10$^{-2}$ \\
     
     SXS:BHNS:0006 & 314 & 6.4$\times$10$^{-3}$ & 7.9$\times$10$^{-3}$ & 7.4$\times$10$^{-3}$ & 7.7$\times$10$^{-3}$ & 6.7$\times$10$^{-3}$ & 7.4$\times$10$^{-2}$ & 1.0$\times$10$^{-2}$ & 1.3$\times$10$^{-2}$ \\
     
     SXS:BHNS:0007 & 315 & 1.2$\times$10$^{-2}$ & 1.9$\times$10$^{-2}$ & 8.1$\times$10$^{-3}$ & 1.3$\times$10$^{-2}$ & 1.1$\times$10$^{-2}$ & 1.8$\times$10$^{-2}$ & 1.2$\times$10$^{-2}$ & 2.2$\times$10$^{-2}$ \\
     
     SXS:BHNS:0008 & 300 & 1.2$\times$10$^{-2}$ & 1.9$\times$10$^{-2}$ & 6.9$\times$10$^{-3}$ & 1.3$\times$10$^{-2}$ & 7.2$\times$10$^{-3}$ & 8.8$\times$10$^{-2}$ & 6.7$\times$10$^{-3}$ & 1.7$\times$10$^{-2}$ \\
     
     SXS:BHNS:0009 & 238 & 2.5$\times$10$^{-2}$ & 4.1$\times$10$^{-2}$ & 7.1$\times$10$^{-3}$ & 1.0$\times$10$^{-2}$ & 7.9$\times$10$^{-3}$ & 8.4$\times$10$^{-3}$ & 8.8$\times$10$^{-3}$ & 2.2$\times$10$^{-2}$ \\ 
      
     \hline\hline
   \end{tabular}
  \label{tab:matching}
 \end{table*}

In addition, for all simulations we compute the unfaithfulness for higher modes of \TEOB{} for different inclinations until merger, employing the same PSD as before. 
These results are obtained using the sky-maximized faithfulness defined in Ref.~\cite{Gamba:2021ydi}. i.e. maximizing over the effective polarizability $\kappa$, in addition to $t_0$ and $\phi_0$, and averaging over the sky localization (we do not need to maximise the faithfulness over the rotations due to precession). 
Results are reported in in Table~\ref{tab:match_HM}. We notice that for $\iota = 0$, the unfaithfulness deviates slightly in comparison to the ones obtained for the (2,2) only waveforms. This is expected with the inclusion of HMs and the use of a different method to compute the unfaithfulness. 
Moreover, the low quality of the few available NR data with HMs has a negative effect on the computed faithfulness, especially for cases such as SXS:BHNS:0005.

 \begin{table}[t]
   \centering    
      \caption{Unfaithfulness of \TEOB{} including HMs with long SXS NR waveforms for different inclinations $\iota$, until merger (top) and for all the waveform (bottom). The results are obtained using the Advanced LIGO zero-detuning, high-power noise curve.}
   \begin{tabular}{c|c|c|c|c}  
     \hline\hline
     \multicolumn{5}{c}{Until merger} \\
     \hline
     Model & $\iota=0$ & $\iota=\pi/8$ & $\iota=\pi/4$ & $\iota=\pi/2$\\
     \hline
     SXS:BHNS:0001 & 9.6$\times$10$^{-3}$ & 1.3$\times$10$^{-2}$ & 2.1$\times$10$^{-2}$ & 3.5$\times$10$^{-2}$ \\
     SXS:BHNS:0002 & 5.9$\times$10$^{-3}$ & 1.2$\times$10$^{-2}$ & 1.7$\times$10$^{-2}$ & 2.6$\times$10$^{-2}$  \\
     SXS:BHNS:0003 & 6.9$\times$10$^{-3}$ & 9.4$\times$10$^{-3}$ & 1.5$\times$10$^{-2}$ & 2.7$\times$10$^{-2}$ \\
     SXS:BHNS:0004 & 1.3$\times$10$^{-2}$ & 1.3$\times$10$^{-2}$ & 1.1$\times$10$^{-2}$ & 1.7$\times$10$^{-2}$   \\
     SXS:BHNS:0005 & 3.8$\times$10$^{-2}$ & 9.7$\times$10$^{-2}$ & 1.6$\times$10$^{-1}$ & 6.4$\times$10$^{-1}$  \\
     SXS:BHNS:0006 & 1.9$\times$10$^{-3}$ & 3.6$\times$10$^{-3}$ & 3.7$\times$10$^{-3}$ & 8.1$\times$10$^{-3}$  \\
     SXS:BHNS:0007 & 1.1$\times$10$^{-2}$ & 1.5$\times$10$^{-2}$ & 2.7$\times$10$^{-2}$ & 4.8$\times$10$^{-2}$ \\
     SXS:BHNS:0008 & 1.7$\times$10$^{-2}$ & 1.3$\times$10$^{-2}$ & 1.6$\times$10$^{-2}$ & 3.3$\times$10$^{-2}$ \\
     SXS:BHNS:0009 & 2.6$\times$10$^{-2}$ & 3.2$\times$10$^{-2}$ & 4.3$\times$10$^{-2}$ & 7.4$\times$10$^{-2}$  \\
     \hline
      \multicolumn{5}{c}{All waveform} \\	
      \hline
     SXS:BHNS:0001 & 9.9$\times$10$^{-3}$ & 1.3$\times$10$^{-2}$ & 2.1$\times$10$^{-2}$ & 3.5$\times$10$^{-2}$ \\
     SXS:BHNS:0002 & 6.4$\times$10$^{-3}$ & 1.4$\times$10$^{-2}$ & 2.2$\times$10$^{-2}$ & 2.6$\times$10$^{-2}$  \\
     SXS:BHNS:0003 & 6.9$\times$10$^{-3}$ & 9.4$\times$10$^{-3}$ & 1.5$\times$10$^{-3}$ & 2.7$\times$10$^{-3}$ \\
     SXS:BHNS:0004 & 1.8$\times$10$^{-2}$ & 1.8$\times$10$^{-2}$ & 1.7$\times$10$^{-2}$ & 1.7$\times$10$^{-2}$   \\
     SXS:BHNS:0005 & 4.2$\times$10$^{-2}$ & 3.2$\times$10$^{-1}$ & 5.3$\times$10$^{-1}$ & 6.3$\times$10$^{-1}$  \\
     SXS:BHNS:0006 & 4.2$\times$10$^{-3}$ & 4.3$\times$10$^{-3}$ & 6.2$\times$10$^{-3}$ & 8.1$\times$10$^{-3}$  \\
     SXS:BHNS:0007 & 1.9$\times$10$^{-2}$ & 2.2$\times$10$^{-2}$ & 4.2$\times$10$^{-2}$ & 4.8$\times$10$^{-2}$ \\
     SXS:BHNS:0008 & 2.5$\times$10$^{-2}$ & 2.5$\times$10$^{-2}$ & 2.8$\times$10$^{-2}$ & 3.3$\times$10$^{-2}$ \\
     SXS:BHNS:0009 & 4.1$\times$10$^{-2}$ & 4.3$\times$10$^{-2}$ & 4.9$\times$10$^{-2}$ & 7.4$\times$10$^{-2}$  \\
     \hline\hline
     \end{tabular}
  \label{tab:match_HM}
 \end{table}

In previous works~\cite{Hinderer:2016eia,Foucart:2018lhe}, and most
recently~\cite{Steinhoff:2021dsn}, BHNS models including the effects
of $f$-mode tidal resonances were developed. In these models the 
fundamental oscillation modes of the NS ($f$-mode) resonates with the 
orbital motion frequency, thus accelerating towards the end of the coalescence. The
works in~\citet{Foucart:2018lhe} and~\citet{Steinhoff:2021dsn} claim
that the contribution on the dynamics is significant enough in low
mass ratio cases to be included in BHNS waveform models. We find that
the phasing and unfaithfulness for the \SEOB{v4T}
model of \citet{Steinhoff:2021dsn} show comparable performance to our
BHNS model in the relevant cases and, therefore, we find no need to
add this effect to our model. Nevertheless the $f$-mode is available in \TEOB{}~\cite{Gamba:2022mgx}. 
\subsection{Precession}
\label{sec:validation:precession}

In the past observing runs of LIGO and Virgo, there have been a
fraction of binaries that exhibit
precession~\cite{LIGOScientific:2020kqk,LIGOScientific:2020ibl,LIGOScientific:2021djp}. This
effect originates when one of the spin vectors is misaligned with
respect to the orbital angular momentum, causing the spin 
and orbital
plane to precess. Based on~\cite{Akcay:2020qrj},~\citet{Gamba:2021ydi}
presented an EOB model for precessing BBH implemented in the newest
version of \TEOB{}
GIOTTO~\cite{Riemenschneider:2021ppj,Gamba:2021ydi}.
This version models quasi-circular precessing and non-precessing BBH, BNS
and (using the model introduced here) BHNS, all including subdominant modes.
To the best of our knowledge, GIOTTO is also the only model incorporating precession effects for BHNS binaries.

\begin{figure*}[t]
  \centering 
  \includegraphics[width=\textwidth]{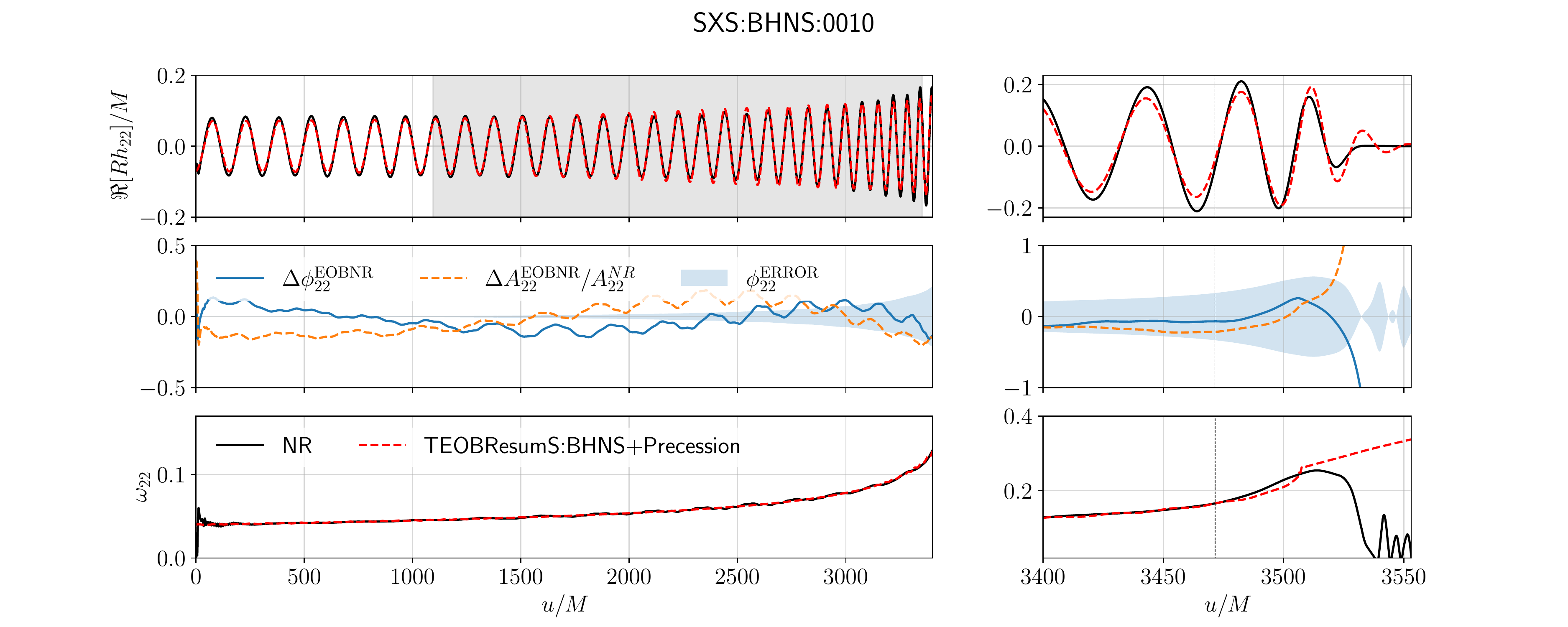}
  \caption{Phasing analysis of the $\ell = m = 2$ mode of SXS:BHNS:0010 against our \TEOB{} model including precession. The plot shows the waveform (top), the phase and amplitude difference with NR (middle), and the frequency (bottom). SXS:BHNS:0010 is the only NR simulation available with a precessing BH spin.}
\label{fig:phasing_prec}
\end{figure*}

Within \TEOB{} GIOTTO one can thus produce precessing BHNS waveforms such as the one shown in Fig.~\ref{fig:phasing_prec}. In this plot we compare the NR simulation SXS:BHNS:0010~\cite{chernoglazov_alexander_2020_4139871,Foucart:2020xkt} with our model. This simulation is currently the only available BHNS NR model with a precessing BH spin. It corresponds to a binary with $q=3$, M$_{\rm{NS}}=1.4\Msun$ and $a_{\rm{BH}}=0.75$ with an initial inclination of 45$^{\circ}$. \TEOB{} shows good agreement through all waveform. The dephasing is comparable to the NR error, especially during and after merger. Specifically, we obtain a phase difference of -0.06 rad and -2.01 rad at merger and in total respectively, with a NR error of -0.33. The combination of the precession and BHNS models allows the EOB waveform to adequately reproduce the NR waveform's morphology without any additional tuning.

\section{Application to GW inference}
\label{sec:inference}
  
In this section we put the model to the test through parameter estimation (PE) on both artificial and available observational data. 
We employ the \bajes{} pipeline~\cite{Breschi:2021wzr} with the {\tt dynesty} nested sampling algorithm~\cite{Speagle:2020}, used to compute posterior probabilities for model parameters and the corresponding Bayesian evidence, see e.g. Refs.~\cite{Veitch:2014wba,LIGOScientific:2019hgc,Breschi:2021wzr} for more details.
In Sec.~\ref{sec:inference:injections} we present simulation studies for two different types of BHNS waveforms. 
A simulated GW signal generated with our model is added on top of Gaussian and stationary noise (this procedure is referred to as ``injection'') generated 
from LIGO and Virgo design sensitivity, and analyzed to obtain the posterior distribution of the binary's parameters. 
Instead, Secs.~\ref{sec:inference:GW190814},~\ref{sec:inference:GW200105} and~\ref{sec:inference:GW200115} report
on PE of GW signals recently observed by the LIGO-Virgo interferometers:
GW190814~\cite{Abbott:2020khf}, GW200105 and GW200115~\cite{Abbott_2021}.
To reduce computational costs, throughout this section we employ the aligned-spin version of the model,
neglecting precession effects. An extension of these studies including precessional degrees of freedom 
will be reported in future work.

Priors used in all inference studies correspond to the range of validity of the model, detailed in Sec.~\ref{sec:validation}. 
We sample on all intrinsic and extrinsic parameters, choosing: flat priors on the component masses, an isotropic prior for aligned spins, and a volumetric prior for the luminosity distance. 
For more details about the functional forms of these and the rest of the parameter's priors, we refer to Sec. V B of~\cite{Breschi:2021wzr}.
For the PE of the GW events, we repeat our studies under different hypotheses: 
BHNS and BBH binary, assuming in turn the presence of the (2,2) mode or HMs for GW190814, and (2,2) mode only for GW200105 and GW200115.
We compare these hypotheses directly through their relative Bayes factors. 
Also, to compare with LVK results, in the analyses of the real events we use the same low spin prior for the NS spin ($a_{\rm{NS}}<$0.05).
Note that in this section we use $D_L$ instead of $R$ to denote the luminosity distance.


\subsection{Injections}
\label{sec:inference:injections}

Two different injected non-precessing signals are
analyzed, with low and high mass ratios and different values of total mass and tidal deformabilities, to account for the different types of BHNS mergers. The injected values of the intrinsic parameters are
reported in Table~\ref{tab:injections}.
Each binary parameter set was injected with 
\texttt{zero noise}
in (2,2) mode only and HMs. Each of these realizations is
recovered with both (2,2) and HMs model, for a consistency test. The injected inclination angle is set to $\iota= 0$ for all cases, while the sky position parameters are set to the location of maximum sensitivity for the \texttt{H1} detector $(\alpha=0.372,\delta=0.811)$.
The luminosity distance value is chosen as to set the SNRs of the injections to the desired values, i.e. 14 and 21.
The injected signals have a sampling rate of 4096 Hz and a length of 64 seconds.
The inference is carried out with 7000 live points within a frequency range of 20 Hz to 2048 Hz. Three detectors are used: \texttt{H1}, \texttt{L1}, and \texttt{V1}, with LIGO-Virgo design sensitivity P1200087~\cite{TheLIGOScientific:2014jea}\cite{TheVirgo:2014hva}. The results are reported in Tab.~\ref{tab:injections} for the quadrupolar injections and Tab.~\ref{tab:injections1_HM} for the ones injected with HMs.

The first injection (I1) corresponds to a low mass ratio BHNS binary, 
for which tidal disruption occurs, i.e. a Type I BHNS, with a SNR of 14.
Injection 2 (I2) is instead a high mass ratio BHNS binary with SNR 21 for which no disruption occurs, corresponding to a Type II BHNS.
As expected, all source parameters are correctly recovered in all injections. The chirp mass is recovered very precisely (with a $0.14\%$ error), and the spin of the BH is always more constrained than the NS one, due to its larger mass, in all cases. Notably, we recover a value for the inclination $\iota$ for the HMs recovery on both injections of I2. However, the injected value does not lie within the $90\%$ credibility interval of our measurement.

For $\Lambda$ the priors
are recovered, which means that the SNRs we consider do 
not allow precise measurements of the tidal polarizability.

\begin{table*}[t]
  \centering   
   \caption{Parameter estimation results for zero noise (2,2)
     mode injections and recoveries with (2,2) mode and HMs. In the table, ``I'' stands for injected values. The values not shown correspond to cases where the prior was recovered.}
   \begin{tabular}{cc|ccc|ccc}  
     \hline\hline
     \multirow{2}{*}{Parameters}  &  & 
     	\multicolumn{3}{c|}{Injection 1} & 
     	\multicolumn{3}{c}{Injection 2} \\
     & & I & (2,2) recovery & HMs recovery & I & (2,2) recovery & HMs recovery \\
     \hline
     $\mathcal{M}$ [$\Msun$] &  & 1.703 & 1.704$^{+0.002}_{-0.001}$  & 1.704$^{+0.002}_{-0.001}$  & 
     2.78  & 2.779$^{+0.005}_{-0.008}$  &  2.778$^{+0.006}_{-0.007}$ \\
     
     $q$ &  & 2 & 2.09$^{+1.04}_{-0.67}$  & 1.70$^{+0.82}_{-0.45}$  & 
     6  & 5.84$^{+0.98}_{-1.38}$  &  5.34$^{+0.95}_{-1.79}$ \\
     
     $m_1$ [$\Msun$] &  & 2.8 & 2.87$^{+0.71}_{-0.53}$  & 2.57$^{+0.61}_{-0.38}$  &  
     8.4 & 8.27$^{+0.78}_{-1.18}$  &  7.86$^{+0.78}_{-1.62}$   \\
     
     $m_2$ [$\Msun$] &  & 1.4 & 1.37$^{+0.27}_{-0.23}$  & 1.51$^{+0.24}_{-0.25}$  &  
     1.4 & 1.42$^{+0.17}_{-0.09}$  & 1.47$^{+0.28}_{-0.10}$    \\
     
     $\abh$ &  & 0 & 0.03$^{+0.17}_{-0.13}$  & -0.01$^{+0.13}_{-0.12}$  &  
     0 & 0.00$^{+0.08}_{-0.15}$  &  -0.05$^{+0.08}_{-0.33}$   \\
     
     $a_{\rm{NS}}$ &  & 0 & -0.01$^{+0.28}_{-0.19}$  & -0.01$^{+0.17}_{-0.14}$  &  
     0 & -0.10$^{+0.22}_{-0.33}$  &  0.02$^{+0.30}_{-0.37}$   \\
      
      $\iota$ [rad] &  & 0 & --  & --  &  
      0 & --  &  0.33$^{+0.27}_{-0.16}$   \\
     
     $D_L$ [Mpc] &  & 400 & 297$^{+157}_{-97}$  & 309$^{+88}_{-99}$  &  
     400 & 292$^{+82}_{-81}$  &  390$^{+26}_{-38}$   \\
     
     $\Lambda$ &  & 791 & --  & --  &  
     526 & --  &  --   \\
     \hline
     SNR &  & 14 & 13.69$^{+0.31}_{-0.47}$  & 13.57$^{+0.57}_{-0.82}$ &  
     21 & 20.75$^{+0.29}_{-0.44}$  &  20.62$^{+0.38}_{-0.91}$    \\
     \hline\hline
   \end{tabular}
  \label{tab:injections}
 \end{table*}
 
 \begin{table*}[t]
  \centering   
   \caption{Same as Table~\ref{tab:injections} but for zero noise HMs injections with (2,2) and HMs recovery.}
   \begin{tabular}{cc|ccc|ccc}  
     \hline\hline
     \multirow{2}{*}{Parameters}  &  & 
     	\multicolumn{3}{c|}{Injection 1} & 
     	\multicolumn{3}{c}{Injection 2} \\
     & & I & (2,2) recovery & HMs recovery & I & (2,2) recovery & HMs recovery \\
     \hline
     $\mathcal{M}$ [$\Msun$] &  & 1.703  & 1.704$^{+0.002}_{-0.001}$ & 1.704$^{+0.002}_{-0.001}$  &  
     2.78  & 2.779$^{+0.006}_{-0.006}$  &  2.781$^{+0.007}_{-0.004}$ \\
     
     $q$ &  & 2 &  2.05$^{+0.98}_{-0.64}$  & 1.96$^{+1.08}_{-0.60}$  &
     6  & 5.56$^{+1.25}_{-1.13}$  &  6.45$^{+2.07}_{-0.71}$ \\

     $m_1$ [$\Msun$] &  & 2.8 &  2.84$^{+0.67}_{-0.51}$  & 2.77$^{+0.75}_{-0.48}$  &   
     8.4 & 8.04$^{+1.00}_{-0.98}$  &   8.76 $^{+1.54}_{-0.57}$  \\
 
     $m_2$ [$\Msun$] &  &  1.4  & 1.38$^{+0.27}_{-0.23}$ & 1.41$^{+0.27}_{-0.26}$  & 
     1.4 & 1.45$^{+0.15}_{-0.12}$ & 1.36 $^{+0.07}_{-0.15}$   \\

     $\abh$ &  & 0 &  0.02$^{+0.16}_{-0.13}$  & 0.02$^{+0.16}_{-0.10}$  &  
     0 & -0.02$^{+0.10}_{-0.14}$  &  0.02$^{+0.17}_{-0.04}$   \\

     $a_{\rm{NS}}$ &  & 0 &  -0.005$^{+0.81}_{-0.36}$  & -0.002$^{+0.376}_{-0.382}$  &  
     0 & 0.11$^{+0.20}_{-0.32}$  & -0.12 $^{+0.30}_{-0.25}$  \\

      $\iota$ [rad] &  & 0 & --  &  --  & 
      0 & --  &  0.34$^{+0.25}_{-0.18}$  \\

     $D_L$ [Mpc] &  & 400 & 288$^{+162}_{-94}$  &  326$^{+73}_{-98}$  & 
     400 & 298$^{+78}_{-86}$  &  387 $^{+30}_{-41}$  \\

     $\Lambda$ &  & 791 & --  & --  &  
     526 & --  &  --   \\
     \hline
     SNR &  & 14 &  13.68$^{+0.32}_{-0.48}$  & 13.58$^{+0.38}_{-0.79}$  & 
     21 & 20.74$^{+0.30}_{-0.41}$  &  20.58 $^{+0.41}_{-0.86}$   \\
     \hline\hline
   \end{tabular}
  \label{tab:injections1_HM}
 \end{table*}

As a consistency check, we compute the Bayes factors between the two recoveries for each injection, as seen in Table~\ref{tab:bayesf}. The results are coherent: in all cases, the Bayes factors favour the recovery done with the corresponding injected modes or show no decisive evidence for the other recovery, thus confirming the model can correctly infer the source parameters. 

\begin{table*}[t]
   \centering    
   \caption{Bayes factors $\log(\mathcal{B})$ for the different
     recoveries done for the HMs injections 1 and 2. The recovery columns represent the different SNRs that were injected in addition to the modes used for recovery. 
     Each entry is computed as the
     difference between the (logarithmic) evidence $\log Z$ of the column
     hypothesis and the row hypothesis, i.e. $\log(\mathcal{B}^{\rm (2,2)~inj})=\log Z^{\rm (2,2)~rec} - \log Z^{\rm HMs~rec}=-2.73\pm0.19$.} 
   \begin{tabular}{c|c||c|c}
     \hline\hline
   	  \multicolumn{4}{c}{Injection 1} \\      
          \hline
          (2,2) injected & (2,2) recovery & HMs injected  & (2,2) recovery\\
          \hline
          HMs recovery& -2.73$\pm$0.19 & HMs recovery & -3.29$\pm$0.19 \\
          \hline\hline
  
     \multicolumn{4}{c}{Injection 2} \\      
          \hline
          (2,2) injected & (2,2) recovery & HMs injected  & (2,2) recovery\\
          \hline
          HMs recovery& 0.36$\pm$0.23 & HMs recovery & -0.19$\pm$0.27 \\
          \hline\hline

   \end{tabular}
   \label{tab:bayesf}
 \end{table*}

\subsection{GW190814}
\label{sec:inference:GW190814}

\begin{figure}[t]
  \centering 
  \includegraphics[width=0.5\textwidth]{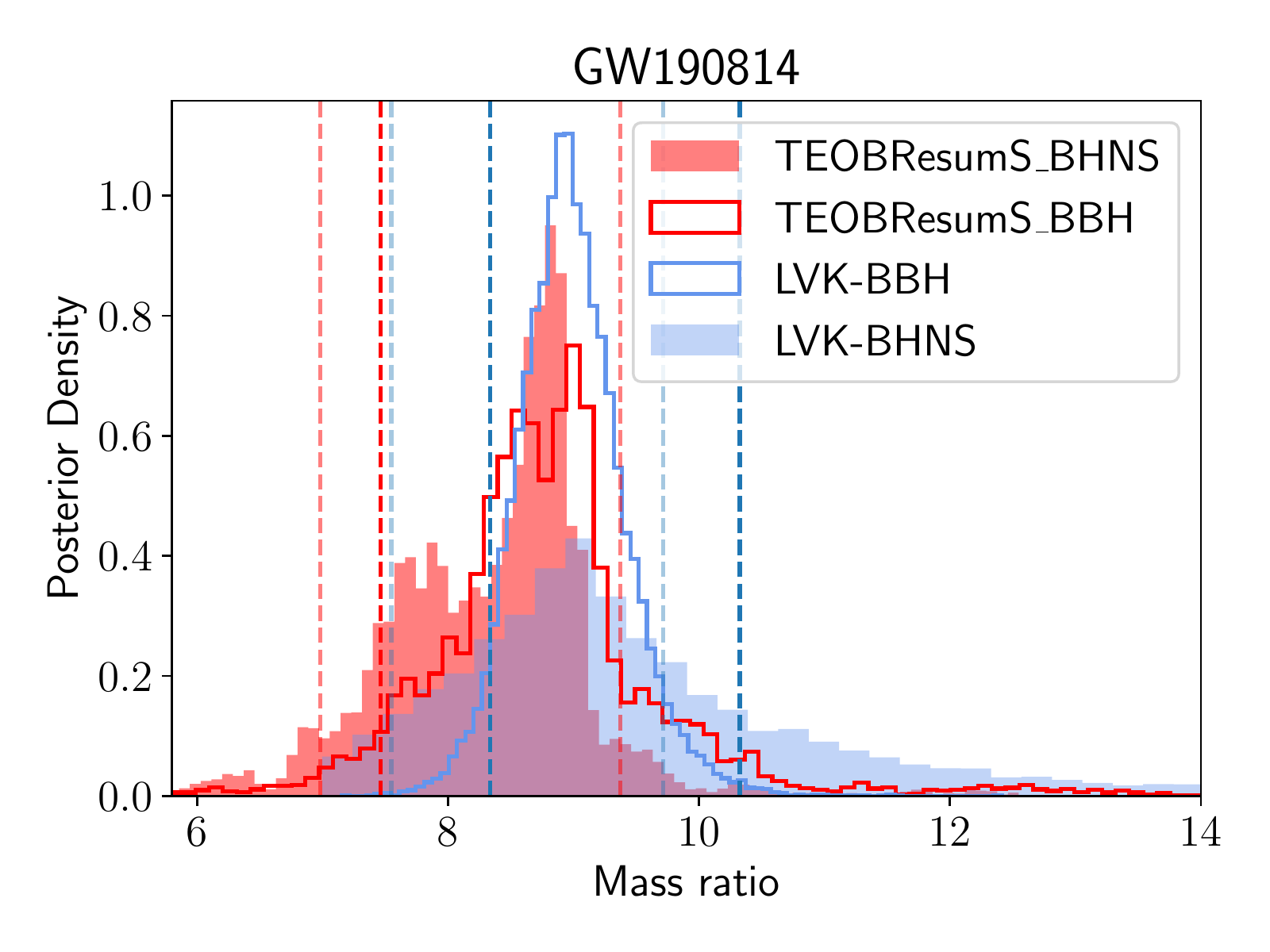}
  \caption{Posterior distribution of the mass ratio $q$. Inference runs were made for two hypothesis (BHNS and BBH) with \TEOB{} and HMs. We add for comparison the publicly available combined posterior from LVK's studies using BBH with HMs models and BHNS with (2,2) mode.}
\label{fig:190814_q}
\end{figure}

We apply our model to the observed signal GW190814. This transient
corresponds to a peculiar GW source, composed of a 23.2 $\Msun$ BH and
a compact object of 2.59 $\Msun$, according to the LVK
analysis~\cite{Abbott:2020khf}. The secondary component of this system
could be either the lightest BH or the heaviest NS ever observed. The
analysis is performed with 2500 live points.
Results can be seen in Table~\ref{tab:PE_GW}, where we compare our BHNS estimated parameters to LVK's BHNS 
data recovered with the \texttt{SEOB} model. 
We find that the our results employing the (2,2) mode BHNS
model are consistent with LVK's ones.
However, the addition of HMs to the analysis recovers a smaller median value for the mass of the primary binary component and a slightly higher value for the secondary component. 
This is reflected in Figure~\ref{fig:190814_q}, which compares the mass ratio posteriors with the LVK result.
In this figure, we also compare to results obtained with BBH models, namely \TEOB{} BBH and LVK's combined BBH, which show a slight preference towards higher mass ratios.
From the recovered parameters in Table~\ref{tab:PE_GW} and from the above mentioned plot, 
one can also observe that, as expected, the addition of HMs to the model helps to further constrain the recovered values. We also note that one can recover a measurement of the inclination in contrast to the (2,2) mode only analysis.
Since the $\Lambda$ posteriors coincide with the prior, no tidal parameters are measured using the BHNS model.
We obtain a slightly higher SNR value when considering HMs, recovering 23.2$^{+0.3}_{-0.4}$ and 24.1$^{+0.5}_{-1.3}$ for (2,2) and HMs respectively. 
These results lie close to the one obtained by LVK with BHNS (2,2) mode model, 23.8$^{+0.1}_{-0.1}$. 
In LVK's analysis, including BBH waveform templates with precession in addition to HMs allows to recover a significantly higher value of 25.0$^{+0.1}_{-0.2}$~\cite{Abbott:2020khf}. We expect future analysis with our BHNS mode including precession (as implemented in the newest \TEOB{} version GIOTTO) to improve the match with the data. 

The \TEOB{} Bayes factors $\mathcal{B}$ for each hypothesis considered for this event the beginning of this section.
In both BHNS and BBH cases, the presence of HMs is strongly favoured compared to the $(2,2)$ only hypothesis.
This finding extends the LVK result, which only derived such preference for the HM hypothesis in the BBH case.
When comparing the BHNS vs. BBH hypotheses, in the $(2,2)$ only case we obtain a weak preference for a BBH origin,
$\log \mathcal{B}_{\mathrm{BBH,(2,2)}}^{\mathrm{BHNS,(2,2)}} = -0.58 \pm 0.21$.
In the HMs case we find $\log \mathcal{B}_{\mathrm{BHNS,HM}}^{\mathrm{BBH,HM}} = 0.13 \pm 0.33$, which means that the BHNS and BBH hypotheses have
comparable evidence, within statistical uncertainty.

\subsection{GW200105}
\label{sec:inference:GW200105}

Next, we analyse the first BHNS observed by the LIGO/Virgo interferometers. 
The analysis is carried out with 4096 live points and considering the dominant mode only. The inclusion of subdominant
modes is left to future work.
Results are shown in Table~\ref{tab:PE_GW}.

\begin{figure}[t]
  \centering   
  \includegraphics[width=0.4\textwidth]{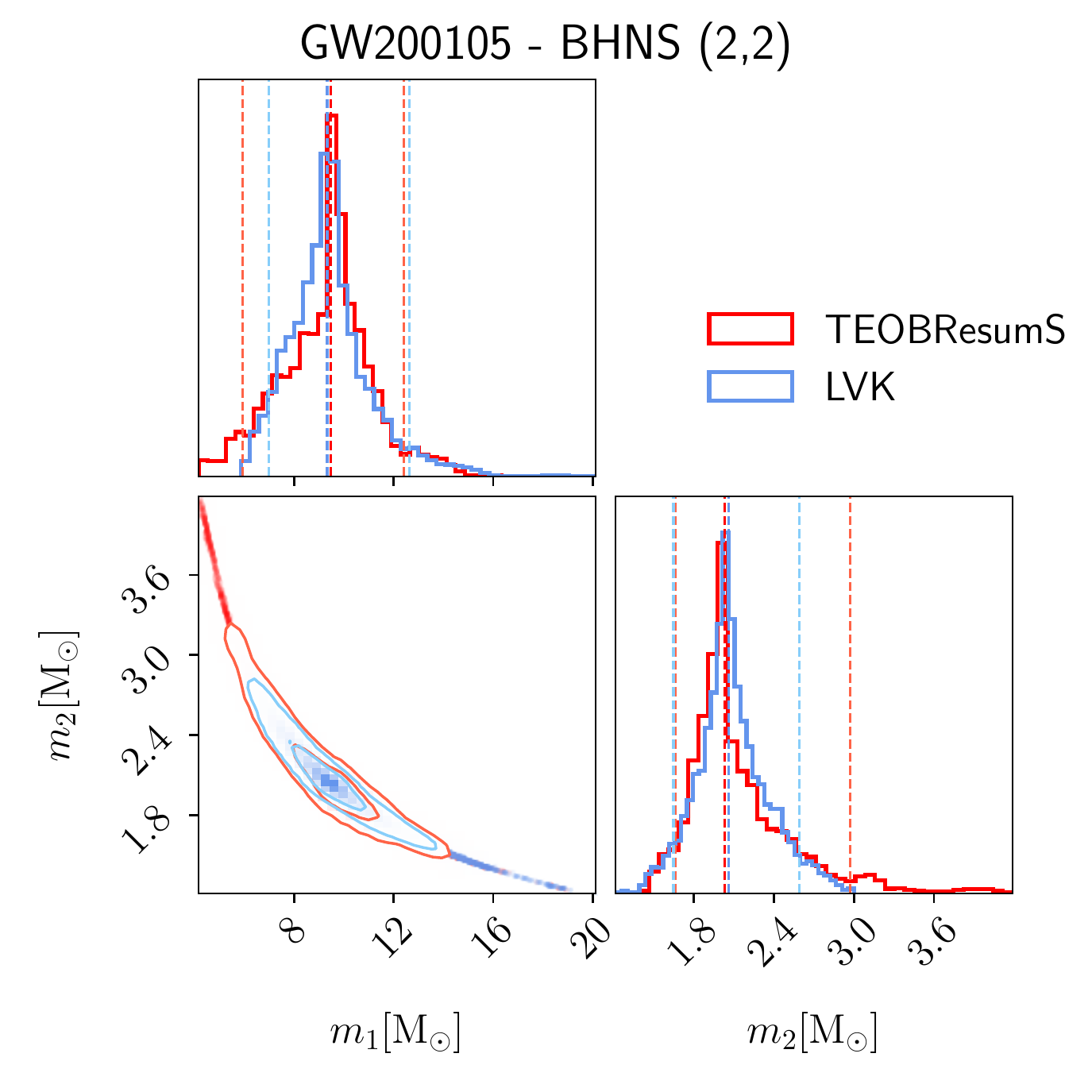}
  \caption{Posterior density for the components' masses of GW200105. 
  Comparison of our results with (2,2) mode together with the LVK results.}
  \label{fig:gw105}
\end{figure}

Our (2,2) mode analysis is broadly consistent with the results obtained by LVK. This can be seen in the 
component masses shown in Figure~\ref{fig:gw105}. We recover again the prior for
$\Lambda$, therefore no tidal parameter measurements are
possible. 

The Bayes factors are reported in Table~\ref{tab:bayesf}. 
The BBH hypothesis is only slightly favoured compared to the BHNS one
$\log\mathcal{B}^{\mathrm{BHNS,(2,2)}}_{\mathrm{BBH,(2,2)}} = -0.19\pm 0.23$. 

Future work, including for instance subdominant modes and precession, may help to better understand the source of this GW event.

\subsection{GW200115}
\label{sec:inference:GW200115}

Finally, we perform PE on GW200115, the second observed BHNS.
As in the previous section, the analysis is carried out using the (2,2) mode only and 4096 live points.
The analysis including HMs will be presented in future work.  
The analysis is carried out similarly as above, with 4096 live points. The results are collected 
in Table~\ref{tab:PE_GW}.

\begin{figure}[t]
  \centering   
  \includegraphics[width=0.4\textwidth]{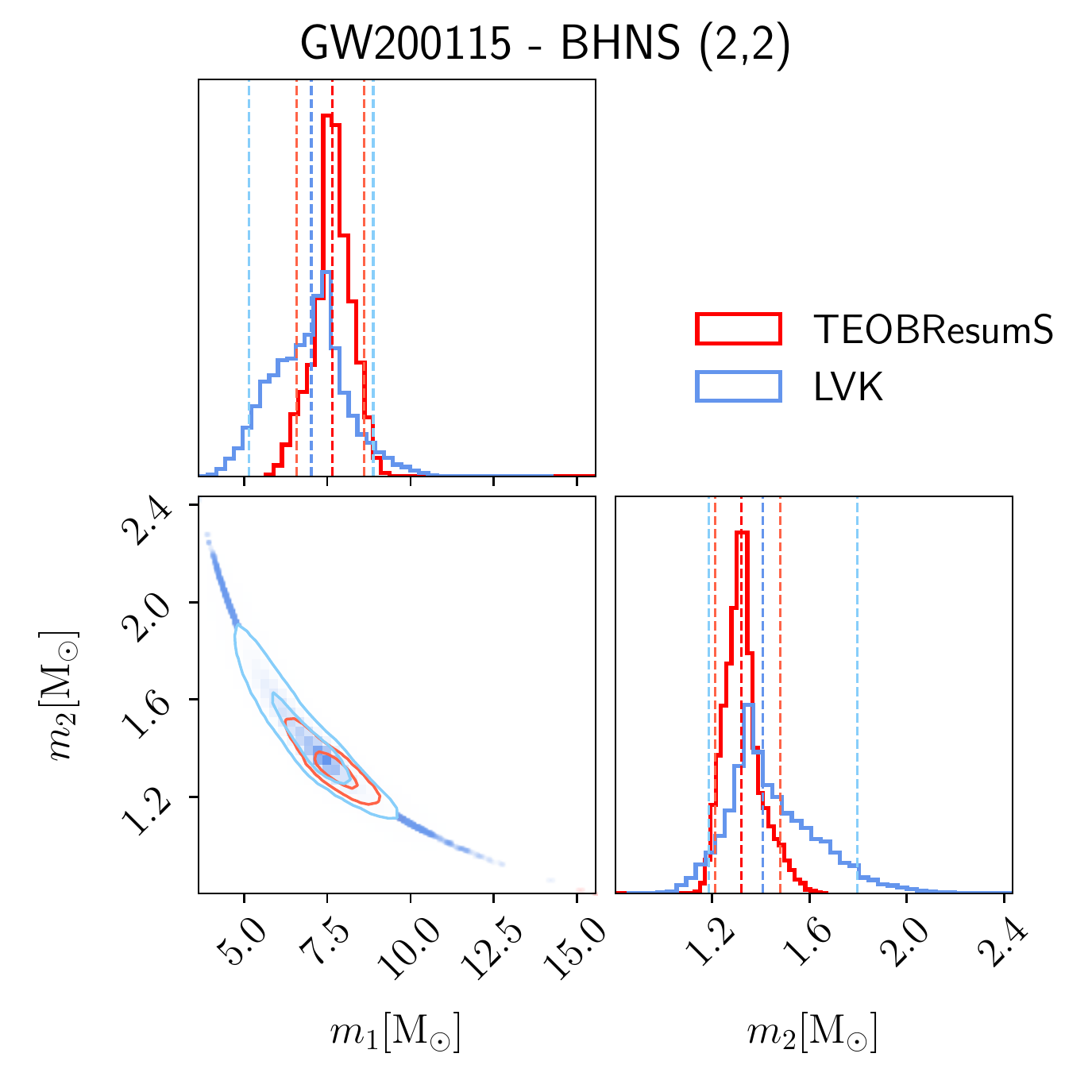}
  \caption{Posterior density for the components' masses of GW200115. The results obtained using the (2,2) mode obtained with our model are compared to the LVK ones.}
  \label{fig:gw115}
\end{figure}

Agreement with LVK results is found within statistical errors.
We find a higher value of the mass ratio 
stemming from the underlying BBH model to which our BHNS model is based~\cite{Damour:2014sva,Nagar:2015xqa,Nagar:2018zoe,Nagar:2019wds,Nagar:2020pcj,Riemenschneider:2021ppj}. 
This implies a slightly smaller value for the NS mass and a more massive BH, as displayed in Fig.~\ref{fig:gw115}. 
The recovered posterior of $\Lambda$ coincides with the prior 
as well for this case.
The Bayes factor comparing the BHNS vs. BBH hypothesis, $\log\mathcal{B}^{\mathrm{BHNS,(2,2)}}_{\mathrm{BBH,(2,2)}} = -2.71\pm
0.25$, indicates no evidence for the presence of tidal effects in the signal.
A similar analysis including the subdominant modes will be presented in future work.

\begin{table*}[t]
  \centering   
   \caption{Parameter estimation of GW events using the BHNS \TEOB{} model, with the dominant $(\ell,m)=(2,2)$ mode and HMs. We compare our results with those obtained by the LIGO-Virgo-Kagra (LVK) Collaboration with BHNS waveform models employing the only available $(\ell,m)=(2,2)$ mode. The values not shown correspond to cases where the prior was recovered.}
   \begin{tabular}{cc|ccc|cc|cc}        
     \hline\hline
     \multirow{2}{*}{Parameters}  &  & 
     	\multicolumn{3}{c|}{GW190814} & 
     	\multicolumn{2}{c|}{GW200105} & 
     	\multicolumn{2}{c}{GW200115}\\
     & & (2,2) & HMs & LVK & (2,2) & LVK & (2,2) & LVK \\
     \hline
     $\mathcal{M}$ [$\Msun$] &  & $6.39^{+0.05}_{-0.04}$ & $6.39^{+0.03}_{-0.02}$ & $6.41^{+0.03}_{-0.02}$ &
      3.62$^{+0.01}_{-0.01}$ &  $3.62^{+0.01}_{-0.01}$&
       2.583$^{+0.004}_{-0.004}$ &   $2.582^{+0.004}_{-0.004}$\\
     
     $q$ &  &$8.98^{+3.57}_{-2.60}$ & $8.50^{+0.44}_{-0.92}$ & $9.18^{+1.83}_{-1.00}$ &
      4.67$^{+1.26}_{-1.69}$ &  $4.53^{+1.37}_{-1.20}$ &
       5.81$^{+0.76}_{-0.72}$ &   $4.99^{+1.21}_{-1.48}$\\
     
     $m_1$ [$\Msun$] &  &$24.37^{+2.27}_{-2.25}$ & $23.61^{+0.71}_{-1.54}$ & $24.73^{+2.88}_{-1.63}$ &
      9.48$^{+1.39}_{-2.12}$ &  $9.32^{+1.53}_{-1.48}$ &
       7.66$^{+0.56}_{-0.56}$ &   $7.02^{+0.94}_{-1.26}$\\
     
     $m_2$ [$\Msun$] &  &$2.71^{+0.19}_{-0.16}$ & $2.78^{+0.13}_{-0.06}$ & $2.69^{+0.13}_{-0.19}$ &
      2.03$^{+0.44}_{-0.20}$ &  $2.06^{+0.30}_{-0.22}$ & 
      1.32$^{+0.08}_{-0.07}$ &  $1.41^{+0.23}_{-0.12}$\\
     
     $\chi_{\rm{eff}}$ &  &-$0.02^{+0.09}_{-0.09}$ & -$0.05^{+0.03}_{-0.06}$ & -$0.01^{+0.10}_{-0.07}$ & 
     0.0$^{+0.1}_{-0.2}$ &  -$0.01^{+0.11}_{-0.13}$ & 
     0.02$^{+0.05}_{-0.06}$ &  -$0.04^{+0.09}_{-0.16}$\\
     
	 $\iota$ [rad] &  & -- & $0.89^{+1.34}_{-0.12}$ & -- & 
     -- &  -- & 
     -- &  -- \\  
     
     $D_L$ [Mpc] &  &$239^{+113}_{-93}$ & $232^{+44}_{-53}$ & $280^{+43}_{-66}$ &
      348$^{+145}_{-117}$ &  $280^{+74}_{-81}$ & 
      500$^{+212}_{-154}$ &   $293^{+90}_{-77}$\\
     
	   $\Lambda$ &  & -- & -- & -- &
      -- &  -- & 
      -- &  -- \\  
     
     \hline
      SNR &  & 23.2$^{+0.3}_{-0.4}$ & 24.1$^{+0.5}_{-1.3}$ & 23.8$^{+0.1}_{-0.1}$ &
       13.2$^{+0.2}_{-0.3}$  &  $13.3^{+0.1}_{-0.2}$ & 
       10.62$^{+0.2}_{-0.4}$ &   $11.0^{+0.2}_{-0.3}$\\
     \hline\hline
   \end{tabular}
  \label{tab:PE_GW}
 \end{table*}


 \begin{table}[t]
   \centering    
   \caption{Bayes factors $\log(\mathcal{B})$ for the different
     PE analysis done. Each entry is computed as the
     difference between the (logarithmic) evidence of the column
     hypothesis and the row hypothesis.} 
   \begin{tabular}{c|cc}
     \hline\hline
   	  \multicolumn{3}{c}{GW190814} \\      
          \hline
          & BHNS (2,2) & BBH HMs \\
          \hline
          BHNS HMs & -23.33$\pm$0.27 & 0.13$\pm$0.33 \\
          BBH (2,2) & -0.58$\pm$0.21 & 22.87$\pm$0.27 \\
          \hline\hline

     \multicolumn{3}{c}{GW200105} \\
     \hline
     & BHNS (2,2)  & BBH HMs \\
     \hline
     BHNS HMs & --  & -- \\
     BBH (2,2) & -0.19$\pm$0.23  & -- \\
     \hline\hline
  
     \multicolumn{3}{c}{GW200115} \\
     \hline
     & BHNS (2,2) & BBH HMs  \\
     \hline 
     BHNS HMs & --  & -- \\
     BBH (2,2) & -2.71$\pm$0.25 & --  \\
     \hline\hline
   \end{tabular}
   \label{tab:bayesf}
 \end{table}

\section{Conclusions} 
\label{sec:conclu}

In this work, we presented the first complete (IMR) model for quasi-circular BHNS waveforms with HMs and precessing spins~\cite{Gamba:2021ydi}. The model extends and completes the \TEOB{} framework for circular compact binaries (GIOTTO).
The main novelties with respect to other models are: 
(i) a new NR-informed remnant BH model that updates the one
from~\citet{Zappa:2019ntl},
(ii) the use of next-to-quasicircular corrections (NQC) to the waveform specifically designed from NR BHNS
simulation data, and
(iii) a NR-informed ringdown model based on~\cite{Damour:2014yha}, suitably ``deformed'' for BHNS.
The waveform model has been informed and tested against 131 NR simulations
available to us, and has been thoroughly tested in the parameter
ranges $q\in[1,20]$, $M_{\rm{NS}}\in[1,3]$, $\Lambda\in[2,5000]$, and
$\abh\in[-0.99,0.99]$. 
The former spin range also applies to the spin of the NS, 
thus expanding the validity range of other available models from $[-0.9,0.9]$ to $[-0.99,0.99]$.  

We developed and validated our model against the NR simulations of
Refs.~\cite{Taniguchi:2007xm,Kyutoku:2010zd,Kyutoku:2011vz,Shibata:2011jka,Foucart:2013psa,Kyutoku:2015gda,Hinderer:2016eia,Foucart:2018lhe,Chakravarti:2018uyi,foucart_francois_2020_4139881,duez_matthew_2020_4139890,Hayashi:2020zmn}. 
This comparison is done through waveform phasings and unfaithfulness computations. 
The phase agreement with NR lies in most cases within numerical error throughout the inspiral, 
with a maximal phase difference of 0.5 rad up to merger. 
Right after merger some larger dephasing might occur, which usually remains within 1 rad. 
These results are reflected in the EOB/NR unfaithfulness. 
The latter shows a good quantitative agreement for the entire waveform between our model and NR. 
Finally, for the first time we present a comparison to SXS:BHNS:0010~\cite{chernoglazov_alexander_2020_4139871,Foucart:2020xkt}, the only available BHNS simulation
including spin-precession, finding an agreement comparable to the spin-aligned case.
Furthermore, we also compare the model with other available approximants. 
With respect to the other models, \TEOB{} shows comparable performance through the inspiral and merger, 
with enhanced NR agreement in specific cases. 
Differently from previous work, we do not find evidence for the need of including 
$f$-mode resonances to faithfully represent the same NR data.

We demonstrated the use of \TEOB{} for BHNS in Bayesian PE
using both artificial (injections) and real data, focusing on non-precessing spins. 
The injections provided validation of the model through target signals of type I and II. 
In addition, the events GW190814, GW200105 and GW200115 were analyzed,
and the results compared against those obtained by LVK, 
which employed BHNS models with the $(2,2)$ mode only. 
We also performed the first PE with BHNS including HMs for GW190814, 
and compared these results to the dominant mode analysis and to their 
equivalent using BBH with or without HMs, through the computation of Bayes factors. 
Due to the large mass ratio of all the considered events, no informative measurement of the NS tidal 
parameter is possible.
The inclusion of HMs considerably tightens parameters constraints;
it will thus be essential in future observing runs to rely on BHNS waveform models including HMs. 
With subdominant modes, it was also possible to measure the inclination of the binary system.
When assuming the dominant mode, our BHNS inferences on all events showed broadly consistent results 
to those obtained by the LVK, within statistical uncertainty. 
Due to its large mass ratio, the analysis of GW190814 reinforced the presence of subdominant modes
even under the BHNS hypothesis (extending an equivalent result by the LVK when using BBH models), 
with a Bayes factor of $\log\B^{\mathrm{BHNS,HM}}_{\mathrm{BHNS,(2,2)}}=23.33\pm0.27$. 

We intend to extend our analyses of these GW events using the BHNS 
model including spin-precession, and plan to report on them in a future publication.

A limitation of this study stems from the availability of a small set of NR 
data~\footnote{This is common to all BHNS waveform models, and constitutes the main reason 
why BHNS templates are less sophisticated than BBH or BNS templates.}.
Model testing would benefit from extra NR waveforms of sufficient lenght 
for a robust alignment ($\gtrsim 10$ orbits), a clean ringdown emission and the inclusion of HMs. 
Such additional data would significantly enhance our model. 
Figure~\ref{fig:2Dfits} and Fig.~\ref{fig:nqc_fits} show that both the ringdown model 
and the NQC parameters would benefit from a finer coverage of the parameter space, 
in particular for mass ratios $q\lesssim 10$ throughout the whole BH spin 
parameter range. %
Finally, exploring large values of the tidal polarizability parameter $\Lambda$ 
is particularly interesting for modelling purposes (though not necessarily from the astrophysical point of view), 
since it would help to better understand and constrain the tidal disruptive mergers.

\begin{acknowledgments}
  A.G., R.G., and M.B. acknowledge support by the Deutsche Forschungsgemeinschaft (DFG) under Grant No. 406116891 within the Research Training Group RTG 2522/1.
  M.B. F.Z., G.C. and SB acknowledge support by the EU H2020 under ERC Starting Grant, no.~BinGraSp-714626.
  GC acknowledges support by the Della Riccia Foundation under an Early Career Scientist Fellowship.
  GC acknowledges funding from the European Union’s Horizon 2020 research and innovation program under the Marie Sklodowska-Curie grant agreement No. 847523 ‘INTERACTIONS’, from the Villum Investigator program supported by VILLUM FONDEN (grant no. 37766) and the DNRF Chair, by the Danish Research Foundation.
  Data analysis was performed on the supercomputers ARA and DRACO at Jena. We acknowledge the computational resources provided
  by Friedrich Schiller University Jena, supported in part by DFG grants
  INST 275/334-1 FUGG and INST 275/363-1 FUGG and EU H2020 under ERC Starting Grant, no.~BinGraSp-714626..
  Data postprocessing was performed on the Virgo ``Tullio'' server 
  in Torino, supported by INFN.

  Data of this paper are available at 

  \url{https://doi.org/10.5281/zenodo.6226935}

  \noindent
  The BHNS model in this work is also implemented in \TEOB{} GIOTTO publicly available at
  
  \url{https://bitbucket.org/eob_ihes/teobresums/}
  
  \noindent
  For reproducibility, the results presented in this paper are done with the ``BHNS" branch.
  
  \noindent
  \bajes{} is publicly available at
  
  \url{https://github.com/matteobreschi/bajes}

  \noindent
  This research has made use of data, software and/or web tools obtained 
  from the Gravitational Wave Open Science Center (\url{https://www.gw-openscience.org}), 
  a service of LIGO Laboratory, the LIGO Scientific Collaboration and the 
  Virgo Collaboration. LIGO is funded by the U.S. National Science Foundation. 
  Virgo is funded by the French Centre National de Recherche Scientifique (CNRS), 
  the Italian Istituto Nazionale della Fisica Nucleare (INFN) and the 
  Dutch Nikhef, with contributions by Polish and Hungarian institutes.
\end{acknowledgments}

\appendix

\section{NR data employed}
\label{app:NRdata}
A list of the NR simulations employed in this work is shown in Table \ref{tab:nrdata}. In addition to these, we use the simulations shown in Table II of~\cite{Zappa:2019ntl}.
\begin{table*}[t]
  \centering    
   \caption{Numerical relativity simulations employed for the ringdown model. The rest of the simulations are described in Table II of~\cite{Zappa:2019ntl}. The SACRA simulations shown in this table were used only for the updated remnant fits of Sec.~\ref{sec:model:remnant}. Simulations SXS:BHNS:0005 and SXS:BHNS:0007 have a spinning NS with $a_{\rm{NS}}=-0.2$.}
   \begin{tabular}{ccccccc}        
     \hline\hline
     Name & Code & $q$ & $M_{\rm NS}$[$M_\odot$] & $a_{\rm BH}$ &
     Modes Avail. & Ref.\\
     \hline
     SXS:BHNS:0001 & SpEC & 6 & 1.4 & 0 & (2,1), (2,2), (3,2), (3,3), (4,4), (5,5) &~\cite{duez_matthew_2019_3311687} \\
     SXS:BHNS:0002 & SpEC & 2 & 1.4 & 0 & (2,1), (2,2), (3,2), (3,3), (4,4), (5,5) &~\cite{kidder_larry_2019_3311707} \\
     SXS:BHNS:0003 & SpEC & 3 & 1.4 & 0 & (2,1), (2,2), (3,2), (3,3), (4,4), (5,5) &~\cite{sxs_collaboration_2019_3311678} \\
     SXS:BHNS:0004 & SpEC & 1 & 1.4 & 0 & (2,1), (2,2), (3,2), (3,3), (4,4), (5,5) &~\cite{kidder_larry_2019_3337271} \\
     SXS:BHNS:0005 & SpEC & 1 & 1.4 & 0 & (2,1), (2,2), (3,2), (3,3), (4,4), (5,5) &~\cite{foucart_francois_2019_3337268} \\
     SXS:BHNS:0006 & SpEC & 1.5 & 1.4 & 0 & (2,1), (2,2), (3,2), (3,3), (4,4), (5,5) &~\cite{buonanno_alessandra_2019_3337269} \\
     SXS:BHNS:0007 & SpEC & 2 & 1.4 & 0 & (2,1), (2,2), (3,2), (3,3), (4,4), (5,5) &~\cite{hinderer_tanja_2019_3337270} \\
     SXS:BHNS:0008 & SpEC & 3 & 1.4 & 0.9 & (2,1), (2,2), (3,2), (3,3), (4,4), (5,5) &~\cite{foucart_francois_2020_4139881} \\
     SXS:BHNS:0009 & SpEC & 4 & 1.4 & 0.9 & (2,1), (2,2), (3,2), (3,3), (4,4), (5,5) &~\cite{duez_matthew_2020_4139890} \\
     125H-Q15 & SACRA & 1.5 & 1.35 & 0 & (2,2) &~\cite{Hayashi:2020zmn}\\
     125H-Q19 & SACRA & 1.9 & 1.35 & 0 & (2,2) &~\cite{Hayashi:2020zmn}\\
     125H-Q22 & SACRA & 2.2 & 1.35 & 0 & (2,2) &~\cite{Hayashi:2020zmn}\\
     125H-Q26 & SACRA & 2.6 & 1.35 & 0 & (2,2) &~\cite{Hayashi:2020zmn}\\
     125H-Q30 & SACRA & 3.0 & 1.35 & 0 & (2,2) &~\cite{Hayashi:2020zmn}\\
     125H-Q37 & SACRA & 3.7 & 1.35 & 0 & (2,2) &~\cite{Hayashi:2020zmn}\\
     125H-Q44 & SACRA & 4.4 & 1.35 & 0 & (2,2) &~\cite{Hayashi:2020zmn}\\
     H-Q15 & SACRA & 1.5 & 1.35 & 0 & (2,2) &~\cite{Hayashi:2020zmn}\\
     H-Q19 & SACRA & 1.9 & 1.35 & 0 & (2,2) &~\cite{Hayashi:2020zmn}\\
     H-Q22 & SACRA & 2.2 & 1.35 & 0 & (2,2) &~\cite{Hayashi:2020zmn}\\
     H-Q26 & SACRA & 2.6 & 1.35 & 0 & (2,2) & \cite{Hayashi:2020zmn}\\
     H-Q30 & SACRA & 3.0 & 1.35 & 0 & (2,2) &~\cite{Hayashi:2020zmn}\\
     H-Q37 & SACRA & 3.7 & 1.35 & 0 & (2,2) &~\cite{Hayashi:2020zmn}\\
     H-Q44 & SACRA & 4.4 & 1.35 & 0 & (2,2) &~\cite{Hayashi:2020zmn}\\
     HB-Q15 & SACRA & 1.5 & 1.35 & 0 & (2,2) &~\cite{Hayashi:2020zmn}\\
     HB-Q19 & SACRA & 1.9 & 1.35 & 0 & (2,2) &~\cite{Hayashi:2020zmn}\\
     HB-Q22 & SACRA & 2.2 & 1.35 & 0 & (2,2) &~\cite{Hayashi:2020zmn}\\
     HB-Q26 & SACRA & 2.6 & 1.35 & 0 & (2,2) &~\cite{Hayashi:2020zmn}\\
     HB-Q30 & SACRA & 3.0 & 1.35 & 0 & (2,2) &~\cite{Hayashi:2020zmn}\\
     \hline\hline
   \end{tabular}
   \label{tab:nrdata}
\end{table*}

\section{Fit models}
\label{app:fitmodel}

\begin{figure*}[t]
  \centering
  \includegraphics[width=0.9\textwidth]{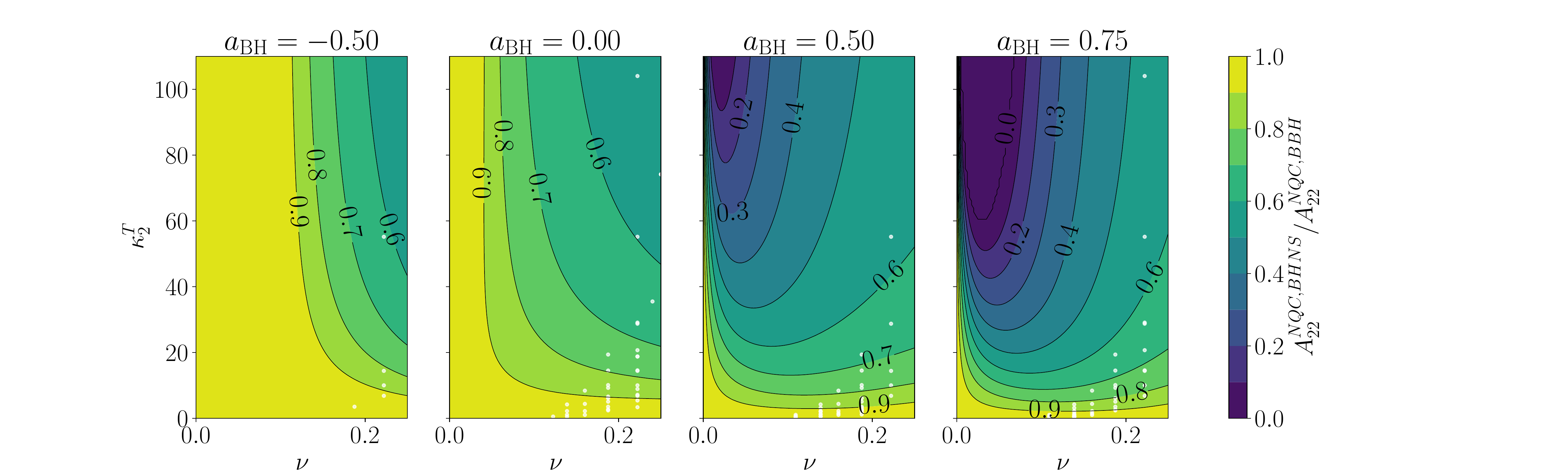}
  \includegraphics[width=0.9\textwidth]{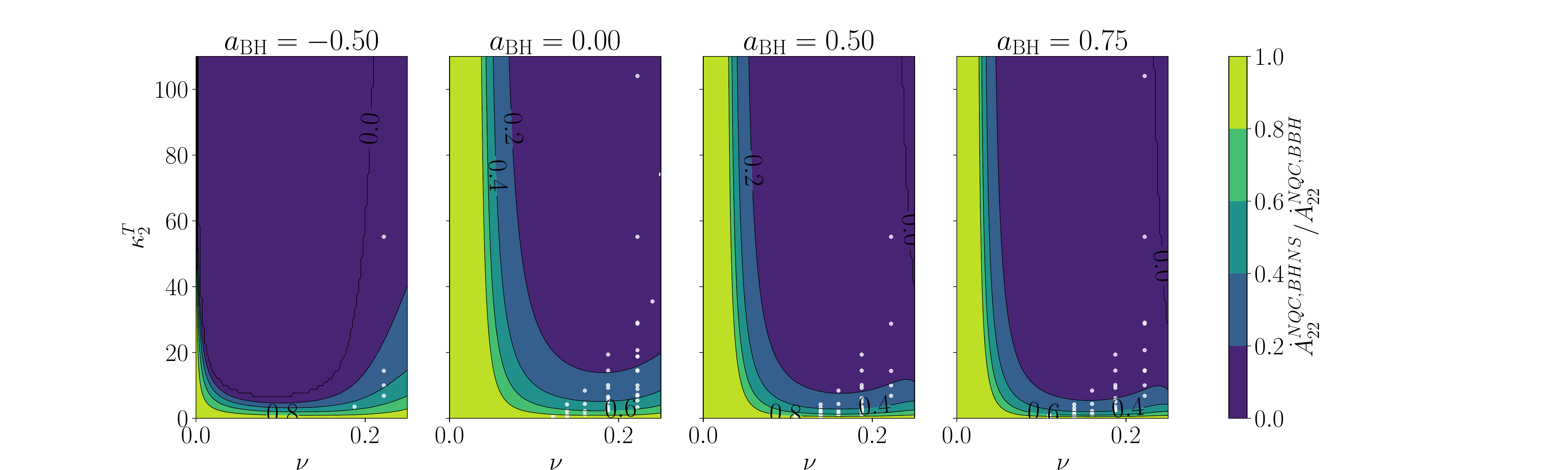}
  \includegraphics[width=0.9\textwidth]{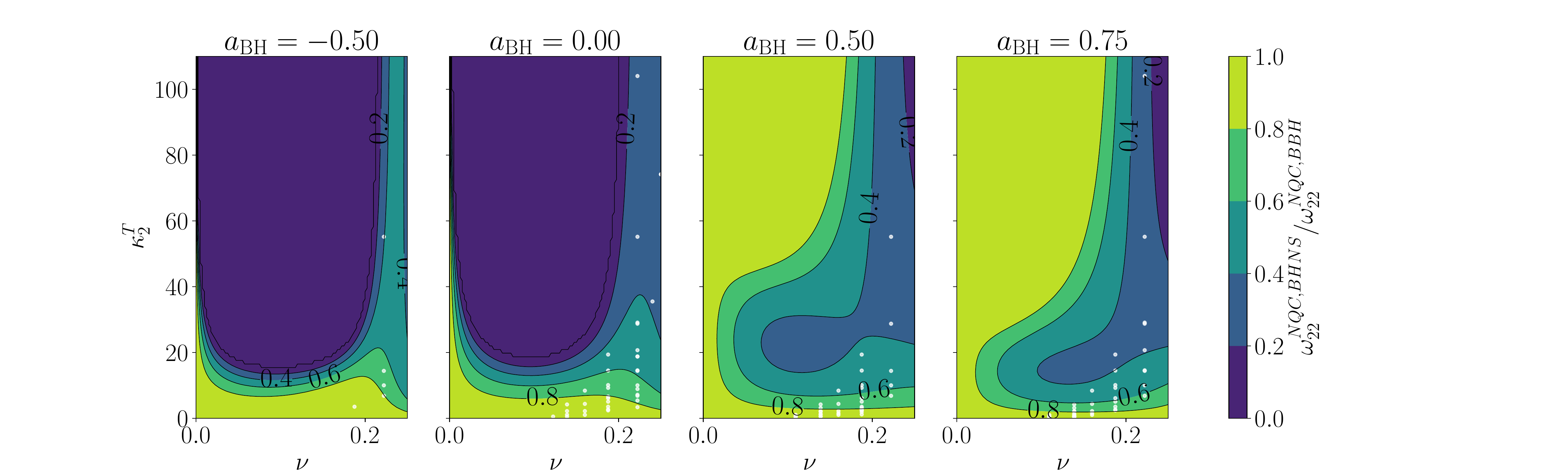}
  \includegraphics[width=0.9\textwidth]{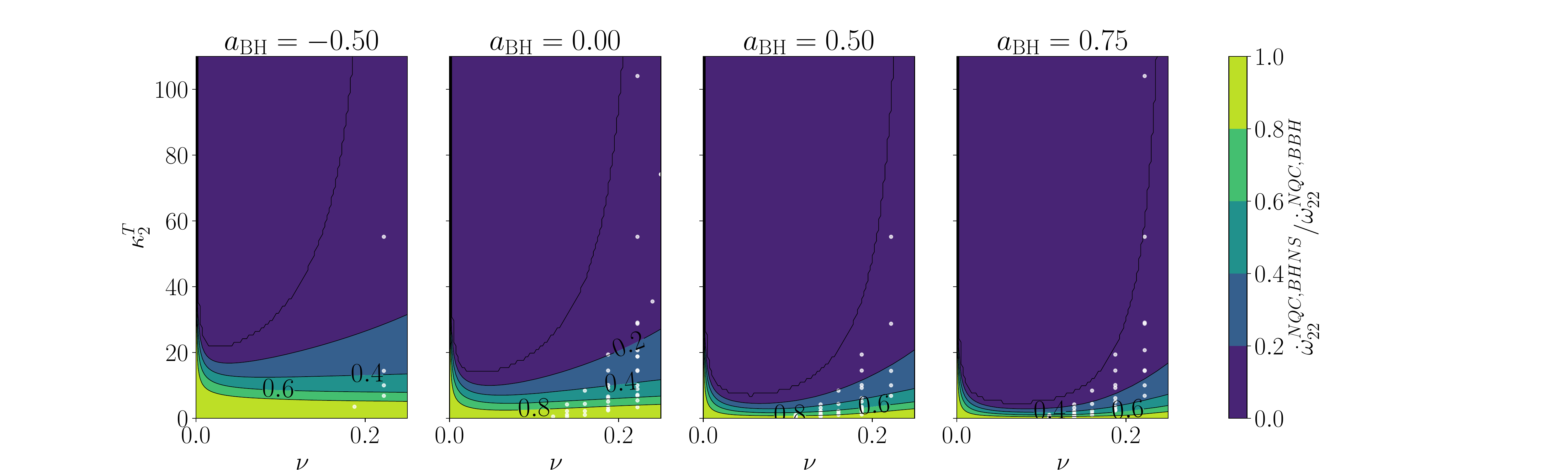}
  \caption{Two-dimensional plots of the NQC quantities' fits: $A_{22},\, \dot{A}_{22},\, \omega_{22},\, \dot{\omega}_{22}$. The fits are developed as functions of ($\nu$,$\kappa^T_2$,$\abh$) (see text). The NR data employed in the fits are represented with white dots.}
  \label{fig:nqc_fits}
\end{figure*}

All the fits developed in this work are based either on Eq.~\eqref{eq:Ffit1} or Eq.~\eqref{eq:Ffit2}. In these equations, $\lambda$ can be either $\kappa^T_2$ or $\Lambda$.
The expression chosen to construct the model, as well as the choice of tidal parameter to use, relies on the performance of the model and how well it captures the numerical results. The performance is measured by computing the coefficient of determination $R^2$.
The BHNS NQC fits were developed in a similar way to the ringdown fits reported in Sec.~\ref{sec:model:ringdown} as described in the main text. The results of the fits can be seen in Fig.~\ref{fig:nqc_fits}, displaying a similar behaviour to the ringdown fits of Fig.~\ref{fig:2Dfits}.
Table~\ref{tab:remnant} shows the updated parameters for the fitting models of the remnant BH's mass $\Mfbh$ and spin $\afbh$. 
The coefficients for the ringdown and NQC quantities are shown in Tables~\ref{tab:fpeaksco} and~\ref{tab:nqcco}, respectively. 

\begin{subequations}
\be\label{eq:Ffit1}
\frac{F(\nu , \lambda, \abh)}{F^{\rm{BBH}}(\nu, \abh)}=
  \frac{1+p_1(\nu , \abh)\lambda + p_2(\nu , \abh)\lambda^2}{(1 + [p_3(\nu, \abh)]^2\lambda)^2},
\ee
where the polynomials $p_k(\nu,\abh)$ are
\bea
p_k (\nu , \abh)&=&p_{k1}(\abh)\nu + p_{k2}(\abh)\nu ^2, \\
p_{kj}(\abh)&=&p_{kj0}\abh+p_{kj1},
\eea
\end{subequations}

\begin{subequations}
\be\label{eq:Ffit2}
\frac{F(\nu , \lambda, \abh)}{F^{\rm{BBH}}(\nu, \abh)}=
  \frac{1+p_1(\nu , \abh)\lambda + p_2(\nu , \abh)\lambda^2}{(1+q(\nu , \abh)\lambda)^2},
\ee
where the polynomials $p_k(\nu,\abh)$ and $q(\nu,\abh)$ are
\bea
p_k (\nu , \abh)&=&p_{k1}(\abh)\nu + p_{k2}(\abh)\nu ^2, \\
p_{kj}(\abh)&=&p_{kj0}\abh+p_{kj1}, \\
q (\nu , \abh)&=&p_{31}\nu,
\eea
\end{subequations}

\begin{table*}[t]
	\centering    
	\caption{Fitting parameters for $\afbh$ and $\Mfbh$ with 
	         $R^2=0.9421$ and $R^2=0.9209$ respectively.
	}
\begin{tabular}{cccccc}        
	\hline\hline
	$F$ & $k$ & $p_{k10}$ & $p_{k11}$ & $p_{k20}$ & $p_{k21}$\\
	\hline
	
	$\afbh$ & 1 & $-5.8066 \times 10^{-3}$ & $8.0478 \times 10^{-3}$ & $2.5417 \times 10^{-2}$ & $2.3044 \times 10^{-2}$\\ 
	& 2 & $-7.8856 \times 10^{-7}$ & $-2.7420 \times 10^{-6}$ & $6.8245 \times 10^{-6}$ & $4.1155 \times 10^{-5}$\\ 
	& 3 & $-2.8968 \times 10^{-2}$ & $2.5204 \times 10^{-1}$ & $1.4410 \times 10^{-1}$ & $-4.2267 \times 10^{-1}$\\
	\hline 
	
	$\Mfbh$ & 1 & $8.2702 \times 10^{-3}$ & $3.0234 \times 10^{-2}$ & $-9.061 \times 10^{-3}$ & $9.4941 \times 10^{-3}$\\ 
	& 2 & $-1.0096 \times 10^{-7}$ & $1.9646 \times 10^{-6}$ & $1.0412 \times 10^{-4}$ & $2.2787 \times 10^{-4}$\\ 
	& 3 & $3.4965 \times 10^{-3}$ & $1.5633 \times 10^{-2}$ & -- & --\\ 
	\hline\hline
\end{tabular}
\label{tab:remnant}
\end{table*} 

\begin{table*}[t]
   \centering    
   \caption{Fitting parameters for $\omega_{2 2 1}$, $\alpha_{2 2 1}$,
     $A^{\rm{peak}}_{22}$ and $\omega^{\rm{peak}}_{22}$.}
   \begin{tabular}{ccccccc}        
     \hline\hline
      $F$ & $k$ & $p_{k10}$ & $p_{k11}$ & $p_{k20}$ & $p_{k21}$ & $R^2$ \\
      \hline 
      
      $\omega_{2 2 1}$ & 1 & -21886.6904 & 32671.7651 & 69276.4427 & -104816.638 \\
      					& 2 & 109.213126 & -73.4308665 & -484.535259 & 373.904119 & 0.96038 \\
      					& 3 & 28.8600443 & -8.14222943 & -126.930553 & 73.1681672 \\
      \hline 
      
      $\alpha_{2 2 1}$ & 1 & 0.08540533 & 0.05952267 & -0.38077744 & -0.20439610 \\
      					& 2 & 9.9329$\times 10^{-6}$ & 4.8199$\times 10^{-5}$ & -2.9158$\times 10^{-5}$ &  -1.9268$\times 10^{-4}$ & 0.94241\\
      					& 3 & 0.21840792 & 0.48995965 & -0.92644561 & -1.14839419 \\
      \hline
      
      $A^{\rm{peak}}_{22}$ & 1 & -0.81310963 & 1.02856956 & 3.04419224 & 1.47499715 \\
      						& 2 & -0.05335408 & 0.01359198  & 0.25160138 & 0.14790936 & 0.96953\\
      						& 3 & 0.02646178 & 0.65802460 & -- & -- \\
      \hline
      
      $\omega^{\rm{peak}}_{22}$ & 1 & -3.70312833 & 2.39550440 & 12.2538726 & -0.80366536 \\
      							 & 2 & 0.02073814 & -0.05079978 & -0.13570448 & 0.50407406 & 0.96712 \\
      							 & 3 & -0.03175737 & 1.04051247 & -- & -- \\
     \hline\hline
   \end{tabular}
  \label{tab:fpeaksco}
\end{table*}

\begin{table*}[t]
   \centering    
   \caption{Fitting parameters for $A^{\rm{NQC}}_{22}$, $\dot{A}^{\rm{NQC}}_{22}$, $\omega^{\rm{NQC}}_{22}$, and $\dot{\omega}^{\rm{NQC}}_{22}$.}
   \begin{tabular}{ccccccc}        
     \hline\hline
      $F$ & $k$ & $p_{k10}$ & $p_{k11}$ & $p_{k20}$ & $p_{k21}$ & $R^2$ \\
      \hline
      $A^{\rm{NQC}}_{22}$ & 1 & -0.76398122 & 0.95760404 & 2.43443187 & 1.11346679 \\
      					& 2 & -0.04517669 & 0.00812775 & 0.19502087 & 0.13476967 & 0.97241 \\
      					& 3 & -0.00976985 & 0.59799088 & -- & -- \\
      \hline 
      
      $\dot{A}^{\rm{NQC}}_{22}$ & 1 & 13.9833426 & 3.82718022 & -58.3833150 & -11.2910418 \\
      					& 2 & 0.20546684 & 0.04017721 & -0.93252149 & -0.19883111 & 0.85271\\
      					& 3 & 8.11227525 & 6.42747644 & -32.2236022 & -18.4991142 \\
      \hline
      
      $\omega^{\rm{NQC}}_{22}$ & 1 & -1.13450258 & -0.52105826 & 4.90257111 & 3.03849913 \\
      						& 2 & 0.11582820 & -0.03543851  & -0.50986874 & 0.16892437 & 0.97336\\
      						& 3 & 2.58632126 & -0.41023648 & -10.8545425 & 6.17597785 \\
      \hline
      
      $\dot{\omega}^{\rm{NQC}}_{22}$ & 1 & -10.1666881 & 3.09188560 & 16.2933259 & 26.7556682 \\
      							 & 2 & 0.04433317 & -0.54327998 & -0.38812308 & 2.25541552 & 0.91437 \\
      							 & 3 & -0.53282251 & 2.63236115 & -- & -- \\
     \hline\hline
   \end{tabular}
  \label{tab:nqcco}
\end{table*}

\section{Phasing of HMs waveforms}
\label{app:HM}

\begin{figure*}[t]
  \centering 
  \includegraphics[width=\textwidth]{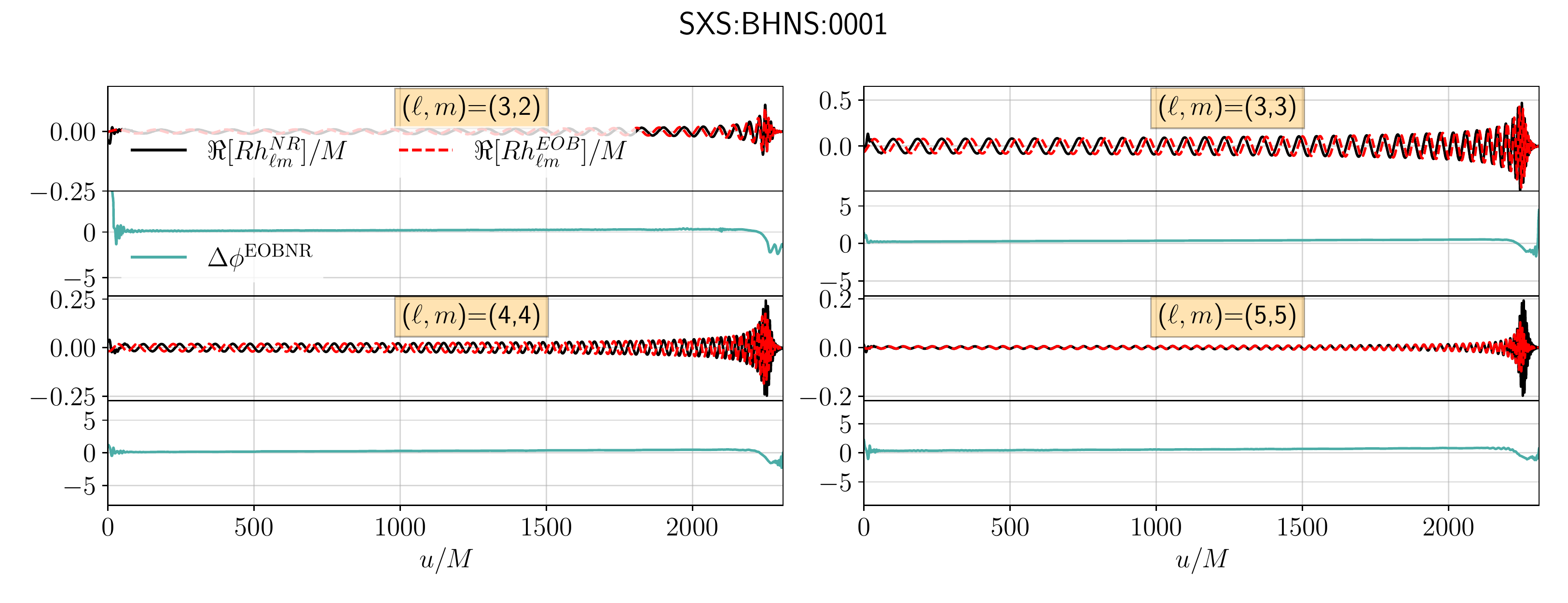}
  \includegraphics[width=\textwidth]{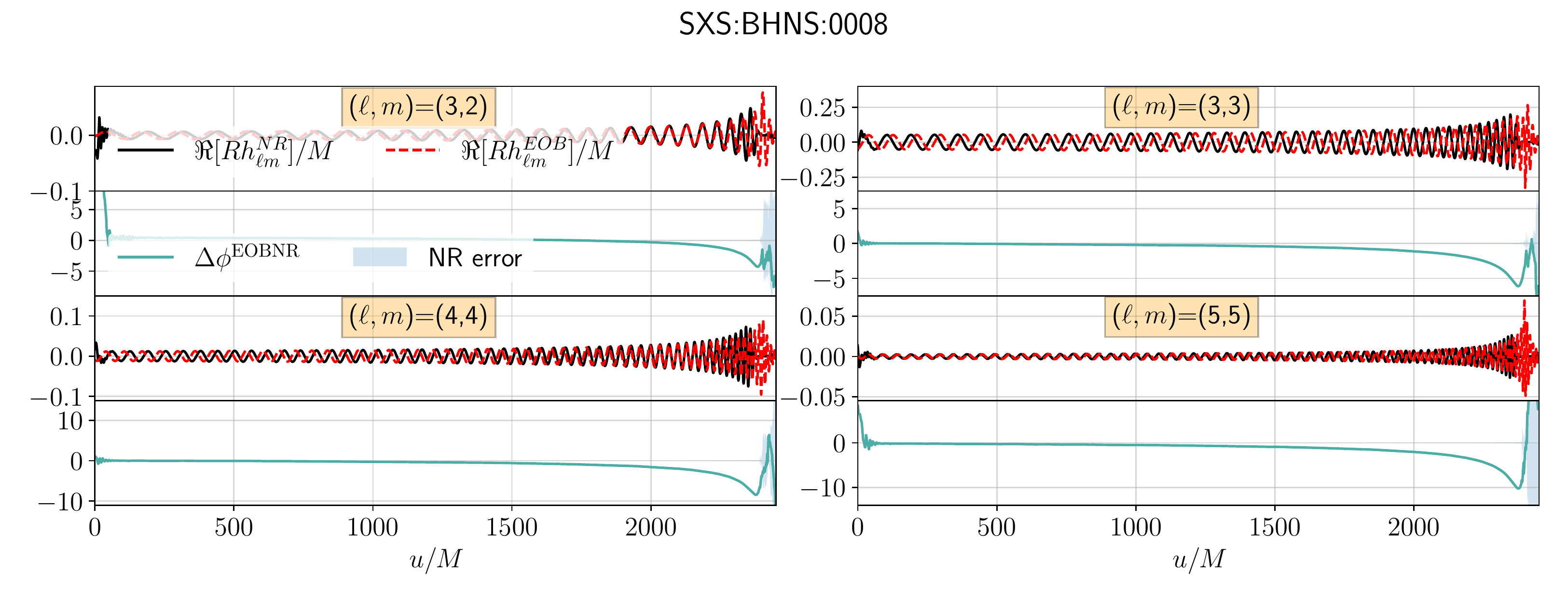}
  \caption{Phasing analysis of SXS:BHNS:0001 (top) and SXS:BHNS:0008 (bottom)
    with \TEOB{}. Light blue bands report the NR resolution error, which is not available for the former.}
\label{fig:phasing_HM}
\end{figure*}

The accuracy of the BHNS HMs waveforms is additionally evaluated through the waveform phasing. Figure~\ref{fig:phasing_HM} shows the alignment of different multipoles from two simulations, SXS:BHNS:0001 (top) and SXS:BHNS:0008 (bottom), against our model. The latter corresponds to a $q=3$ binary with a high spinning BH, $a_{\rm{BH}}=0.9$. Our model shows good agreement through the inspiral and starts dephasing towards and after merger. The phase difference stays however approximately within numerical error after merger for the spinning case.

\clearpage

\end{document}